\documentstyle[aps,prc,preprint]{revtex}
\begin {document}
\draft
\preprint{SUNY-NTG-97-04}
\title
{Rho Meson Propagation and Dilepton Enhancement in Hot Hadronic Matter}
 
\author
{R. Rapp$^{1}$, G. Chanfray$^2$ and J. Wambach$^{3,4}$}
 
\address
{1) Department of Physics, State University of New York, 
    Stony Brook, NY 11794-3800, U.S.A.\\
 2) IPN-Lyon, 43 Bd. de 11 Nov. 1918, F-69622 Villeurbanne Cedex,
    France \\
 3) Institut f\"ur Kernphysik, TH Darmstadt, Schlo{\ss}gartenstr.9,
    D-64289 Darmstadt, Germany \\
 4) Department of Physics, University of Illinois, Urbana,
    IL 61801, U.S.A.}
 
\maketitle
 
\begin{abstract}
A realistic model for the free rho meson with coupling to 
two-pion states is employed to calculate the rho propagator in 
a hot and dense hadron gas. The medium modifications are based 
on hadronic rescattering processes: intermediate two-pion states 
are renormalized through interactions with surrounding nucleons 
and deltas, and rho meson scattering is considered off nucleons, 
deltas, pions and kaons. Constraints from gauge invariance as well 
as the full off-shell dynamics of the interactions are accounted for. 
Within the vector dominance model we apply the resulting in-medium 
rho spectral function to compute $e^+e^-$ production rates 
from $\pi^+\pi^-$ annihilation. The calculation of corresponding 
$e^+e^-$ spectra as recently measured in central collisions of 
heavy-ions at CERN/SpS energies  
gives reasonable agreement with the experimental data. 
\end{abstract}


%%%%%%%%%%%%%%%%%%%%%%%%%%%%%%%%%%%%%%%%%%%%%%%%%%%%%%%%%%%%%%%%%%%%%%%%
\section{Introduction}
%%%%%%%%%%%%%%%%%%%%%%%%%%%%%%%%%%%%%%%%%%%%%%%%%%%%%%%%%%%%%%%%%%%%%%%%
The main objective of ultrarelativistic heavy-ion collisions (URHIC's)
is to explore the phase diagram of strongly interacting hadronic 
matter, in particular possible transitions associated with 
deconfinement and/or chiral symmetry restoration. Due to it's 
negligible rescattering within the hadronic medium, electromagnetic 
radiation (photons and dileptons) is believed to be the most suitable 
probe to study the highest excitation states formed in the early stages
of central collisions. In this context, recent measurements of dilepton 
invariant mass 
spectra~\cite{CERES,DrUl,HELIOS3} in high-energy reactions of  protons 
and sulfur nuclei with heavy target nuclei have raised considerable 
theoretical discussion. In comparison to the p-A case, central S-A 
collisions show a strong increase in the dilepton yield 
(after normalizing to the total number of charged particles). 
As several hundreds of pions are produced in the sulfur induced 
reactions, this increase has been attributed to $\pi^+\pi^- \to 
l^+l^-$ annihilation. However, when including this process in 
numerical transport simulations~\cite{CEK,LKB} (or hydrodynamical 
approaches~\cite{Shur,Prak}) of the collision dynamics, the 
experimentally observed enhancement in the invariant mass range 
$M_{l^+l^-}\simeq 0.3-0.6$~GeV/c$^2$ still remains unexplained.\\ 
Several mechanisms that might increase the dilepton yield below the 
$\rho$ mass have been proposed. In particular, the assumption of a 
a temperature and density dependent dropping $\rho$ 
mass~\cite{BR91} as a precursor of the chiral phase transition  
has been shown to give good agreement with both the CERES and HELIOS-3 
dilepton data~\cite{CEK,LKB}. However, before one can identify this 
mechanism as an unambiguous signal of (partial) chiral symmetry 
restoration, the consequences of more 'conventional' scattering  
processes in the hadronic gas have to be well under control. \\ 
In this article we will address this issue, thereby focussing on 
medium modifications of the $\rho$ meson that are due to 
phenomenologically rather well established hadronic interactions. 
Such processes typically generate a broadening of the $\rho$ spectral 
function and therefore also increase 
the dilepton production at invariant masses below the $\rho$ mass.\\  
The coupling of the $\rho$ meson to intermediate 2-pion states in 
connection with a modified pion propagation in hot/dense matter has 
been analyzed {\it e.g.} in refs.~\cite{HeFN,AKLQ,ChSc,CRW,SKLK}. 
In particular, the interaction of the pions with surrounding nucleons 
and (thermally excited) $\Delta$'s accumulates substantial        
strength in the $\rho$ spectral function for invariant masses below 
0.6~GeV/c$^2$. Further in-medium scattering processes of the $\rho$ 
meson have also been shown to work in the same direction, as {\it e.g} 
resonant $\rho$-nucleon scattering~\cite{FrPi} or interactions with 
pseudoscalar mesons~\cite{Hagl}. 
We will here extend on our previous analysis~\cite{CRW} by including 
such processes in a consistent off-shell treatment  
of $\rho$ propagation in hot hadronic matter. 
The inclusion of the proper off-shell dynamics is of particular 
importance in the low invariant mass region, where the experimentally 
observed excess of dilepton pairs is most pronounced. \\  
Our article is organized as follows: in the next sect. we first 
introduce our model for the $\rho$ meson in free space, which agrees 
with experimental data on p-wave $\pi\pi$ scattering and the pion
electromagnetic form factor in the relevant (timelike) kinematic
regime. We then briefly review the medium modifications of the 
$\rho$ meson generated by a renormalization of the pion propagation
in a $\pi$$N$$\Delta$ gas as discussed in ref.~\cite{ChSc,CRW}. In the
third sect. we turn to the evaluation of in-medium $\rho$-baryon 
scattering. In addition to $\rho N \to N(1720), \Delta(1905)$ 
contributions first discussed by Friman and Pirner~\cite{FrPi}, we
also account for $\rho$-like $NN^{-1}$, $\Delta N^{-1}$, $N\Delta^{-1}$,
and $\Delta\Delta^{-1}$ excitations. In sect.~4 we calculate the 
$\rho$ selfenergy arising from $\rho\pi$ and $\rho K$/$\rho \bar K$ 
scattering. Within the vector dominance model (VDM) the full 
in-medium $\rho$ propagator is employed to calculate $e^+e^-$ production
rates in sect.~5. Making use of recent transport results to model
the temperature/density evolution of central 200~GeV/u S-Au and 
158~GeV/u Pb-Au collisions, and including experimental acceptance, we 
finally compare our results to the CERES $e^+e^-$ data.   
 
%%%%%%%%%%%%%%%%%%%%%%%%%%%%%%%%%%%%%%%%%%%%%%%%%%%%%%%%%%%%%%%%%%%%%%%%
\section{Free $\rho$ Mesons and Medium Modifications in 
$\pi\pi$ Propagation} 
%%%%%%%%%%%%%%%%%%%%%%%%%%%%%%%%%%%%%%%%%%%%%%%%%%%%%%%%%%%%%%%%%%%%%%%%
\subsection{The Model for the $\rho$ Propagator in the Vacuum} 
%%%%%%%%%%%%%%%%%%%%%%%%%%%%%%%%%%%%%%%%%%%%%%%%%%%%%%%%%%%%%%%%%%%%%%%%
Our model for $\rho$ propagator in free space consists of a bare $\rho$ 
meson with mass $m_\rho^{bare}$ renormalized by coupling to intermediate
2-pion states. The basic $\rho\pi\pi$ interaction vertex is 
described by the standard form 
\begin{equation} 
{\cal L}_{\rho\pi\pi} = g_{\rho\pi\pi} (\pmb{$\pi$} \times \partial^\mu
\pmb{$\pi$} ) \cdot \pmb{$\rho$}_\mu \ ,  
\label{Lrhopipi} 
\end{equation} 
where $\pmb{$\pi$}$ and $\pmb{$\rho$}_\mu$ denote the 
isovector pion field
and vector-isovector rho meson field, respectively. 
After resummation of the $\pi\pi$ bubbles to all orders the scalar 
part of the $\rho$ propagator can be cast in the form  
\begin{equation} 
D_\rho^0(M)=[M^2-(m_\rho^{bare})^2-\Sigma_{\rho\pi\pi}^0(M)]^{-1} \ , 
\label{drho0}
\end{equation} 
which in free space depends on the invariant mass $M^2=q_0^2-{\vec q}^2$
only. The $\rho$ meson selfenergy due to coupling to 2-pion states 
is calculated as 
\begin{eqnarray} 
\Sigma_{\rho\pi\pi}^0(M) & = & \bar{\Sigma}_{\rho\pi\pi}^0(M)
-\bar{\Sigma}_{\rho\pi\pi}^0(0) \ ,
 \nonumber\\
\bar{\Sigma}_{\rho\pi\pi}^0(M) & = & \int \frac{k^2 dk}{(2\pi)^2} \ 
v_{\rho\pi\pi}(k)^2 \ G_{\pi\pi}^0(M,k)  \  ,  
\label{sigrho0}
\end{eqnarray}
with the vacuum 2-pion propagator
\begin{equation}
G_{\pi\pi}^0(M,k)=\frac{1}{\omega_k^\pi} \  
\frac{1}{M^2-(2\omega_k^\pi)^2+i\eta}
 \ ; \quad \omega_k^\pi=\sqrt{m_\pi^2+k^2} \ 
\end{equation}
and the vertex functions 
\begin{equation} 
v(k) = \sqrt{\frac{2}{3}} \ g_{\rho\pi\pi} \ 2k \ F_{\rho\pi\pi}(k) \ .
\label{vrhopipi} 
\end{equation} 
The hadronic (dipole) form factor 
\begin{equation}
F_{\rho\pi\pi}(k)=\biggl (\frac{2\Lambda_\rho^2+m_\rho^2}
{2\Lambda_\rho^2 +4\omega_k^2} \biggr)^2 \ 
\end{equation}
accounts for the finite size of the $\rho\pi\pi$ vertex and is 
normalized to one at the physical resonance energy $m_\rho$=0.77~GeV. 
The subtraction of the $\rho$ selfenergy at zero energy is necessary 
to ensure the correct normalization $F_\pi(0)$=1 for the pion 
electromagnetic form factor, which in the vector dominance model (VDM) 
is given by 
\begin{eqnarray}
|F_\pi^0(M)|^2 & = & \frac{(m_\rho^{bare})^4}{(M^2-(m_\rho^{bare})^2-
Re\Sigma_\rho^0(M))^2+(Im\Sigma_\rho^0(M))^2} \nonumber\\
& \equiv & (m_\rho^{bare})^4 \ |D_\rho^0(M)|^2 \ .
\end{eqnarray}
Choosing $g_{\rho\pi\pi}^2/4\pi$=2.7, $\Lambda_\rho$=3.1~GeV and 
$m_\rho^{bare}$=0.829~GeV yields a satisfactory description of the 
vacuum $\pi\pi$ p-wave scattering phase shifts and the pion
electromagnetic form factor in the timelike region (cp. fig.~1). 
 
%%%%%%%%%%%%%%%%%%%%%%%%%%%%%%%%%%%%%%%%%%%%%%%%%%%%%%%%%%%%%%%%%%%%%%%%
\subsection{$\pi\pi$ Propagation in a Hot $\pi$N$\Delta$ Gas} 
%%%%%%%%%%%%%%%%%%%%%%%%%%%%%%%%%%%%%%%%%%%%%%%%%%%%%%%%%%%%%%%%%%%%%%%%
In general, the $\rho$ meson propagator in hot/dense matter does not
solely depend on invariant mass. Since the notion of 
temperature specifies a certain rest frame of the heat bath, Lorentz
invariance is broken and the vector meson propagator splits into 
longitudinal and transverse modes (see {\it e.g.} ref.~\cite{GaKa}): 
\begin{equation} 
D_\rho^{\mu\nu}(q_0,\vec q) = \frac{P_L^{\mu\nu}}{M^2-(m_\rho^{bare})^2
-\Sigma_\rho^L(q_0,\vec q)}+\frac{P_T^{\mu\nu}}{M^2-(m_\rho^{bare})^2
-\Sigma_\rho^T(q_0,\vec q)}+\frac{q^\mu q^\nu}{(m_\rho^{bare})^2 M^2} 
\label{drhomunu} 
\end{equation} 
with the standard projection operators
\begin{eqnarray} 
P_L^{\mu\nu} & = & \frac{q^\mu q^\nu}{M^2}-g^{\mu\nu}-P_T^{\mu\nu} 
\nonumber\\ 
P_T^{\mu\nu} & = &  \left\{ \begin{array}{l}
 \quad~~ 0 \qquad , \mu=0 \ {\rm or} \ \nu=0 
 \\
\delta^{ij}-\frac{q^iq^j}{{\vec q}^2} \ , \ \mu,\nu \in \lbrace 1,2,3 
\rbrace
\end{array}   \right . \ .
\label{PLT}
\end{eqnarray}
(the spacelike components of $\mu$ and $\nu$ will be denoted by 
$i$ and $j$, respectively). 
The longitudinal and transverse selfenergies are defined by the 
corresponding decomposition of the polarization tensor:
\begin{equation} 
\Sigma_\rho^{\mu\nu}(q_0,\vec q)= \Sigma_\rho^L(q_0,\vec q) \ 
P_L^{\mu\nu} + \Sigma_\rho^T(q_0,\vec q) \ P_T^{\mu\nu} \ . 
\end{equation} 

To calculate medium modifications of the $\rho$ meson which are 
generated by dressing the intermediate 2-pion states, we  
adopt the approach of Chanfray and Schuck~\cite{ChSc}, extended to
finite temperature in ref.~\cite{CRW}. 
%A detailed derivation of the corresponding in-medium $\rho$ selfenergy 
%can be found in ref.~\cite{ChSc}. 
Restricting ourselves to the case of back-to-back kinematics, the 
$\rho$ meson selfenergy tensor can be written as 
\begin{equation}
\Sigma_{\rho\pi\pi}^{ij}(q_0,\vec q=\vec 0)=\delta^{ij} \ 
\Sigma_{\rho\pi\pi}(q_0,\vec q =\vec 0) \ . 
\label{sgrppij}
\end{equation} 
Including constraints from gauge invariance the imaginary part of the 
scalar $\rho$ selfenergy at given temperature $T$ and nucleon-delta 
density $\rho_{N\Delta}$=$\rho_N$+$\rho_\Delta$ takes the 
form~\cite{ChSc}  
\begin{eqnarray} 
{\rm Im} \Sigma_{\rho\pi\pi}(q_0,\vec 0) & = & - 
\int\limits_{0}^{\infty} \frac{k^2 \ dk}{(2\pi)^2} 
\ v_{\rho\pi\pi}(k)^2 \int\limits_0^{q_0} \frac{dk_0}{\pi}
\ [1+f^\pi(k_0)+f^\pi(q_0-k_0)] \ {\rm Im} D_\pi(q_0-k_0,k)
\nonumber\\
 & &   * \ {\rm Im}  \lbrace
\alpha(q_0,k_0,k) D_\pi(k_0,k)+\frac{1}{2k^2} \Pi_L(k_0,k)+
\frac{1}{k^2} \Pi_T(k_0,k) \rbrace . 
\label{imsgrpp}
\end{eqnarray}
with the longitudinal and transverse spin-isospin response functions
\begin{eqnarray} 
\Pi_L(k_0,k) & = & (k_0^2-(\omega_k^\pi)^2) \ \tilde\Pi^0(k_0,k)
\ D_L(k_0,k) \ ,
\nonumber\\
\Pi_T(k_0,k) & = & (k_0^2-(\omega_k^\rho)^2) \ \tilde\Pi^0(k_0,k)
\ D_T(k_0,k) \ ,
\end{eqnarray} 
respectively.  The corresponding longitudinal and transverse 
propagators read 
\begin{eqnarray}
D_L(k_0,k) & = & \lbrack k_0^2-(\omega_k^\pi)^2-
\Sigma_\pi(k_0,k) \rbrack ^{-1} \ ,  \nonumber\\
D_T(k_0,k) & = & \lbrack k_0^2-(\omega_k^\rho)^2-
C_\rho\Sigma_\pi(k_0,k) \rbrack ^{-1} \ , 
\end{eqnarray}
with $C_\rho$=2.2 and 
\begin{equation}
\tilde\Pi^0(k_0,k)  =  \frac{1}{k^2} \Sigma_\pi(k_0,k) \ , 
\end{equation} 
where $\Sigma_\pi(k_0,k)$ denotes the in-medium single-pion selfenergy, 
{\it i.e.} $D_L$ is equal to the single-pion propagator $D_\pi$.    
The real function $\alpha$ characterizes corrections to the 
$\pi\pi\rho$-vertex and is related to $\tilde\Pi^0_R(k_0,k) \equiv 
{\rm Re} \tilde\Pi^0(k_0,k)$ by  
\begin{equation} 
\alpha(q_0,k_0,k)= 1+\tilde\Pi^0_R(k_0,k)+\tilde\Pi^0_R(q_0-k_0,k)
+\frac{1}{2}\tilde\Pi^0_R(k_0,k) \tilde\Pi^0_R(q_0-k_0,k) \ . 
\end{equation}
The bosefactors $f^\pi(k_0)=[\exp (k_0/T)-1]^{-1}$ in 
eq.~(\ref{imsgrpp}) account for the thermal 
occupation of the pions in the intermediate state. 
The real part of the $\rho$ 
selfenergy is obtained from a dispersion integral: 
\begin{equation}
{\rm Re} \Sigma_{\rho\pi\pi}(q_0)=-{\cal P} \int\limits_0^\infty
\frac{dE'^2}{\pi} \frac{{\rm Im} \Sigma_{\rho\pi\pi}(E')}{q_0^2-E'^2}
\frac{q_0^2}{E'^2} \ .   
\label{resgrpp}
\end{equation}
The factor ${q_0^2}/{E'^2}$ arises from a  subtraction at zero energy 
which ensures gauge invariance in the limit of vanishing 
three-momentum $\vec q$. \\
The single-pion selfenergy $\Sigma_\pi$ is evaluated within the 
well-known model of particle-hole excitations~\cite{ErWe} extended to 
finite temperature~\cite{RW94}. Here the pion interacts with 
surrounding nucleons and thermally excited $\Delta$'s through
excitations of the type  $NN^{-1}$, $\Delta N^{-1}$, 
$N\Delta^{-1}$ and $\Delta\Delta^{-1}$. The corresponding thermal 
(retarded) Lindhard function in a given excitation channel 
$\alpha$=$ab^{-1}$, $a$,$b$=$N$,$\Delta$, reads 
\begin{equation} 
\phi_\alpha(\omega,k)=-\int \frac{p^2 dp} {(2\pi)^2} 
f^b(E^b_p) \int\limits_{-1}^{+1}
 \ dx \ \sum_{m=1}^{2} \ \frac{1-f^a(E_{pk}^a(x))}
 {\pm\omega+E_p^b-E_{pk}^a(x) \pm
 \frac{i}{2}(\Gamma_a+\Gamma_b)}   \
\label{Lindhard1} 
\end{equation}
including the direct (m=1, signe '+') and the exchange (m=2, 
signe '--') diagram with   
\begin{eqnarray}
E_p^{N,\Delta} & = & (m_{N,\Delta}^2+p^2)^{1/2} \ , \nonumber\\ 
E_{pk}^{N,\Delta}(x) & = & (m_{N,\Delta}^2+p^2+k^2+2pkx)^{1/2} \ . 
\end{eqnarray} 
For the $\Delta$ width $\Gamma_{\Delta}$ we choose a relativistic 
parametrization given in Ref.~\cite{ErWe}, but neglect any in-medium 
width for nucleons, {\it i.e.} $\Gamma_N\equiv 0$. 
The thermal Fermi distributions, 
\begin{equation}
f^a(E_p^a)= (1+\exp[(E_{p}^a-\mu_a)/T])^{-1} \ , 
\label{fermi} 
\end{equation}
determine the nucleon and $\Delta$ densities at given temperature 
$T$ and chemical potential $\mu_N$, $\mu_\Delta$ to be 
\begin{eqnarray}
\rho_N(T) & = & 4 \ \int\frac{d^3q}{(2\pi)^3} \
f^N(E_q^N;\mu_N,T) \nonumber\\
\rho_\Delta(T) & = & 16  \int\frac{d^3q}{(2\pi)^3} \
f^\Delta(E_q^\Delta;\mu_\Delta,T) \ .
\end{eqnarray}
Throughout this article we will assume chemical equilibrium 
characterized by a common baryon chemical potential 
$\mu_B$$\equiv$$\mu_N$=$\mu_\Delta$ and $\mu_\pi$=0. \\ 
%It is worthwhile to mention here that an equivalent result for the 
%thermal Lindhard functions is obtained within the imaginary time 
%(Matsubara) formalism (see {\it e.g.} ref.~\cite{NeOr}): employing 
%standard techniques for performing the Matsubara summations and after 
%analytic continuation one arrives at 
%\begin{equation} 
%\phi_\alpha(\omega,k)=-\int \frac{p^2 dp}{(2\pi)^2}  
%\int\limits_{-1}^{+1} \ dx \  \frac{f^b(E^b_p)-f^a(E_{pk}^a(x))}
%{\omega+E_p^b-E_{pk}^a(x)+\frac{i}{2}(\Gamma_a+\Gamma_b)}  \ , 
%\label{Lindhard2}
%\end{equation}
%which, in the limit of zero temperature, coincides with 
%eq.~(\ref{Lindhard1}) for $\alpha = NN^{-1}, \Delta\Delta^{-1}$ and 
%the sum of $\Delta N^{-1}$ and $N\Delta^{-1}$ excitations. \\  
From the  Lindhard functions one obtains the so-called pionic 
susceptibilities $\chi_\alpha^{(0)}$ as~\cite{ErWe} 
\begin{equation} 
\chi_\alpha^{(0)}(\omega,k)  =  {\left
( \frac{f_{\pi\alpha} \  F_{\pi\alpha}(k)} {m_\pi}
\right )}^2  \ SI(\alpha) \ \phi_\alpha(\omega,k) \ ,
\label{chi0}
\end{equation} 
with spin-isospin factors $SI(\alpha)$, a standard monopole 
form factor 
\begin{equation} 
F_{\pi\alpha}(k)= \left( \frac{\Lambda_\pi^2-m_\pi^2}{\Lambda_\pi^2
+k^2} \right) 
\end{equation}
($\Lambda_\pi$=1.2~GeV) and coupling constants $f_{\pi\alpha}$  
related via $f_{\pi N\Delta}$=2$f_{\pi NN}$ (Chew-Low 
factor~\cite{ChLo}) and  $f_{\pi\Delta\Delta}$=$\frac{1}{5} f_{\pi NN}$ 
(constituent quark model estimate~\cite{BrWe}), cp. table~\ref{tab1}. 
For a realistic description of the pion 
selfenergy the bare susceptibilities have to be corrected for 
short-range correlation effects between particle and hole. These 
are parametrized in terms of Migdal parameters 
${g'}_{\alpha\beta}$, leading to a system of coupled equations  
\begin{equation}
\chi_\alpha  =  \chi_\alpha^{(0)}-\sum_{\beta} \
\chi_\alpha^{(0)} \ {g'}_{\alpha\beta} \ \chi_\beta \ . 
\label{chi}
\end{equation} 
It is solved by an elementary matrix inversion with the final 
result~\cite{RW94}
\begin{equation}
\Sigma_{\pi}(\omega,k)= -k^2 \ \sum_\alpha \chi_{\alpha}(\omega,k)  \ .
\end{equation} 
The Migdal parameters are fixed to ${g'}_{\alpha\beta}=0.8$ for 
$\alpha\beta$=$aa^{-1}bb^{-1}$ and ${g'}_{\alpha\beta}=0.5$ for 
all others. 

\noindent 
With the diagonal form eq.~(\ref{sgrppij}) of the $\rho$ selfenergy 
tensor the corresponding in-medium propagator can be written as
\begin{equation} 
D_\rho^{ij}(q_0,\vec q=\vec 0) = \delta^{ij} \ 
D_\rho(q_0,\vec q=\vec 0) \ ,
\label{drhoij} 
\end{equation} 
where, in analogy to the free case eq.~(\ref{drho0}), the scalar part
is simply
\begin{equation} 
D_\rho(q_0,\vec q=\vec 0)=[M^2-(m_\rho^{bare})^2-
\Sigma_{\rho\pi\pi}(q_0,\vec q=\vec 0)]^{-1} \ . 
\label{drhopipi}
\end{equation} 
In fig.~2, real and imaginary part of $D_\rho$ are displayed for 
different temperatures $T$ at a fixed baryon chemical potential of
$\mu_N$=$\mu_\Delta$=0.39~GeV (this value is chosen with respect to  
what will be used for calculating dilepton spectra in central S-Au 
collisions in sect.~5). With increasing temperature (which at fixed 
$\mu_N$, $\mu_\Delta$  corresponds to a simultaneous 
 increase in $\rho_{N\Delta}$=$\rho_N$+$\rho_\Delta$) 
we find a strong broadening of the $\rho$ spectral function. Although 
the peak is moderately shifted to higher energies (in accordance with 
the findings of refs.~\cite{HeFN,AKLQ,ChSc}), an appreciable 
enhancement over the vacuum result is observed for invariant masses 
below $M$=0.6~GeV. \\  
In the calculation of $e^+e^-$ production rates from in-medium 
$\pi^+\pi^-$ annihilation (sect.~5) the $\rho$ spectral function will 
be weighted with an additional factor of $1/M^2$ steming from an 
intermediate photon propagator. It is instructive to see how this 
combines with our results for Im~$D_\rho$ (cp. fig.~3). 
 At highest temperatures the major part of 
the strength resides in the invariant 
mass range $M$=0.1--0.5~GeV, which 
simply reflects the softening of the pion dispersion relation: the 
low-lying $NN^{-1}$ and $\Delta\Delta^{-1}$ modes lead to a population 
of invariant masses around $M$=0.2~GeV, whereas the $\Delta N^{-1}$ 
excitations, strongly washed out by thermal motion and the large 
$\Delta$ width, show up at somewhat higher $M$$\simeq$0.4~GeV.    
In addition, the shift of strength to low 
invariant masses is accelerated
by the appearance of the pion bosefactors in eq.~(\ref{imsgrpp}).
Even at lower densities one clearly recognizes various combinations of 
pionic branches in the $1/M^2$-weighted $\rho$ spectral function. \\
To investigate the role that the pionic collectivity plays for the 
build-up of the medium effects just discussed, we replaced the 
full propagators $D_\pi$=$D_L$ and $D_T$, entering eq.~(\ref{imsgrpp}), 
by their vacuum forms and recalculated the corresponding $\rho$ 
propagator, cp. fig.~4: without the pion collectivity the broadening  
of the $\rho$ spectral function is strongly reduced, and the 
peak is now slightly shifted downwards (short-dashed line). For small 
temperatures this trend becomes even more pronounced (dotted line).  
A similar behavior has recently been found in a calculation to 
lowest order in the nuclear density~\cite{KlWe}. 

%%%%%%%%%%%%%%%%%%%%%%%%%%%%%%%%%%%%%%%%%%%%%%%%%%%%%%%%%%%%%%%%%%%%%%%%
\section{Rho Scattering in Baryonic Matter}
%%%%%%%%%%%%%%%%%%%%%%%%%%%%%%%%%%%%%%%%%%%%%%%%%%%%%%%%%%%%%%%%%%%%%%%%
Besides the medium modifications caused by dressing the intermediate 
2-pion states, direct interactions of the $\rho$ meson with 
surrounding hadrons in the gas have to be considered. In analogy to 
the in-medium behavior of pions one might expect 
strong effects from $\rho$ scattering off baryons (provided their 
abundance in the hadronic gas is appreciable as seems the case 
in central URHIC's at CERN energies~\cite{Sorge}). From meson 
exchange models such as the Bonn potential~\cite{MHE} one certainly  
knows that {\it e.g.} the $\rho NN$ (or $\rho N\Delta$) coupling 
constant is quite sizeable, even though the corresponding s-channel 
process $\rho N \to N \to \rho N$ is  
kinematically strongly disfavored. Such a kinematic suppression will 
be much less pronounced with increasing energy 
of the resonance in the intermediate state.  
Indeed, there are at least two
well established resonances in the particle data table~\cite{PDG} 
which strongly couple to the $\rho N$ decay channel, namely 
$N$(1720) with isospin/spin-parity $I(J^P)$=$\frac{1}{2}(\frac{3}{2}^+)$
and a branching ratio 
$\Gamma_{N(1720)\to\rho N}/\Gamma_{N(1720)}^{tot}>0.7$,
and $\Delta$(1905) with $I(J^P)$=$\frac{3}{2}(\frac{5}{2}^+)$, 
$\Gamma_{\Delta(1905)\to\rho N}/\Gamma_{\Delta(1905)}^{tot}>0.6$. 
This led Friman and Pirner to the 
idea~\cite{FrPi} of considering $\rho$-like particle-hole excitations 
('rhosobars') in nuclear matter of the type $\rho N(1720)N^{-1}$ and 
$\rho\Delta(1905)N^{-1}$. In analogy to the well-known 'pisobar' 
($\pi\Delta N^{-1}$), the (non-relativistic) interaction Lagrangian
for the two 'rhosobars' can be written as  
\begin{eqnarray} 
{\cal L}_{\rho N^*N} & = & \frac{f_{\rho N^*N}}{m_\rho} \ 
\Psi^\dagger_{N^*} \ ({\vec S}_\frac{3}{2} \times {\vec q}) \cdot 
( \pmb{$\tau$} \cdot \vec{\pmb{$\rho$}}) \ \Psi_N \ + \ {\rm} h.c.  
\nonumber\\
{\cal L}_{\rho \Delta^*N} & = & \frac{f_{\rho \Delta^*N}}{m_\rho} \ 
\Psi^\dagger_{\Delta^*} \ {S}_\frac{5}{2}^{ij} \ q^i \    
({\bf T} \cdot {\pmb{$\rho$}}^j) \ \Psi_N \  + \ {\rm} h.c.
\label{Lrhosobar}
\end{eqnarray} 
($N^*\equiv N(1720)$, $\Delta^*\equiv\Delta(1905)$). Here, 
${\vec S}_{\frac{3}{2}}$ (${S}_\frac{5}{2}^{ij}$) denote the spin 
$\frac{1}{2} \to \frac{3}{2}$ $(\frac{5}{2})$ transition 
operator (tensor), $\pmb{$\tau$}$, $\pmb{$T$}$  the isospin 
$\frac{1}{2}$ and the isospin 
$\frac{1}{2}\to\frac{3}{2}$ transition operator, respectively,
 and ${\vec q}$ the 3-momentum of the $\rho$ meson. 
The coupling constants $f_{\rho B^*N}$ ($B^*$=$N^*$ or $\Delta^*$) are 
adjusted to reproduce the partial decay widths $\Gamma_{B^*\to\rho N}$. 
To  obtain realistic values for them it is important to
account for the finite width of $\rho$ meson, since both resonances 
(especially the $N(1720)$) lie rather close to the $\rho N$ threshold 
(the assumption of a sharp $\rho$ meson would 
allow for only a small phase
space, which in turn would result in an overestimate of the coupling 
constant). Starting from the general expression for the differential 
two-body decay width, given {\it e.g.} in ref.~\cite{ItZu}, we arrive at
\begin{equation} 
\Gamma_{B^*\to \rho N}(\sqrt{s})= \frac{f_{\rho B^*N}^2}{4\pi m_\rho^2}
 \   \frac{2m_N}{\sqrt{s}} \ \overline{SI}(B^*\to\rho N) 
\int\limits_{2m_\pi}^{M_{max}} 
\frac{M dM}{\pi} \ A_\rho(M) \ q_{cm}^3 \ F_{\rho B^* N}(q_{cm})^2 
\label{gammaB*} 
\end{equation}   
where $\overline{SI}(N^*\to\rho N)$=2, 
$\overline{SI}(\Delta^*\to\rho N)$=1/3 are initial-state averaged, 
final-state summed spin-isospin factors and $F_{\rho B^* N}(q_{cm})$ 
is a hadronic monopole form factor ($\Lambda_{\rho B^*N}$=2~GeV) 
depending on the $\rho$/$N$ decay 3-momentum 
in the resonance rest frame,  
\begin{equation}
q_{cm}^2=\frac{(s-M^2-m_N^2)^2-4m_N^2M^2}{4s} \ .
\end{equation}
The $\rho$ meson spectral function, 
\begin{equation} 
A_\rho(M)=-2 \ {\rm Im} D_\rho(M) \ , 
\end{equation} 
is taken from eq.~(\ref{drho0}) and the upper integration limit is
$M_{max}$=$\sqrt{s}-m_N$. With a branching ratio of 70\% for the
$N(1720)$ decay into $\rho N$ and a total width of 0.15~GeV at the 
resonance energy $m_{N^*}$=1.72~GeV, eq.~(\ref{gammaB*}) determines
the coupling to be $f_{\rho N^* N}^2/4\pi$=6.99. Applying the same 
procedure for the $\Delta(1905)$ with a branching ratio to $\rho N$ 
of 60\% we find $f_{\rho\Delta^* N}^2/4\pi$=10.64. \\
Having fixed the coupling constant we can now compute the in-medium 
selfenergy tensors for $\rho$-like $B^*N^{-1}$ excitations from
the interaction vertices of eq.~(\ref{Lrhosobar}), leading to 
\begin{eqnarray} 
\Sigma_{\rho N^*N^{-1}}^{(0),ij}(q_0,\vec q) & = & \left( 
\frac{f_{\rho N^* N} \ F_{\rho N^* N}(q)}{m_\rho}\right)^2 \ 
IF(\rho N^* N^{-1}) \ ~ \phi_{\rho N^*N^{-1}}(q_0,\vec q) 
\nonumber\\   
 & & * \sum_{\lambda_N,\lambda_{N^*}} 
\langle \frac{3}{2}\lambda_{N^*}|({\vec S}_{\frac{3}{2}}
\times{\vec q})^i |\frac{1}{2} \lambda_N\rangle \  
\langle \frac{1}{2}\lambda_{N}|({\vec S}_{\frac{3}{2}}^\dagger\times
{\vec q})^j |\frac{3}{2} \lambda_{N^*}\rangle  
\end{eqnarray} 
and a similar  expression for the $\rho \Delta^* N^{-1}$ bubble 
involving the ${S}_{\frac{5}{2}}$ tensor. 
Both the isospin factors 
\begin{eqnarray} 
IF(\rho N^* N^{-1}) & = & 2 
\nonumber\\
IF(\rho \Delta^* N^{-1}) & = & \frac{4}{3} 
\end{eqnarray} 
and the Lindhard functions 
\begin{equation}
\phi_{\rho\alpha}(q_0,\vec q)=-\int \frac{p^2 dp}{(2\pi)^2} 
f^N(E^N_p) \int\limits_{-1}^{+1}
 \ dx \ \sum_{m=1}^{2} \ \frac{1-f^{B^*}(E_{pq}^{B^*}(x))}
 {\pm q_0+E_p^N-E_{pq}^{B^*}(x) \pm
 \frac{i}{2}\Gamma_{B^*}^{tot}}   
\label{Lindhardrho}
\end{equation}
($\alpha=N^*N^{-1},\Delta^*N^{-1}$) are equivalent to the pionic case, 
eq.~(\ref{Lindhard1}). The total widths are each taken as the sum of 
$\rho N$ channel and $\pi N$ channel, 
\begin{equation}
\Gamma_{B^*}^{tot}=\Gamma_{B^*\to \rho N}+\Gamma_{B^*\to \pi N} \ ,
\end{equation} 
where the $\pi B^* N$ coupling constant is chosen such that the total 
width matches it's experimental value at the resonance mass. \\
The transversality of the $\rho N^*N$ coupling automatically ensures 
that the vector current is conserved, {\it i.e.} 
\begin{equation}
q_i \ \Sigma_{\rho N^*N^{-1}}^{(0),ij} = 0 \ , 
\end{equation} 
which, when coupling to the photon, is a necessary condition for 
gauge invariance. Thus, for the $\Delta(1905)$ contribution we 
only take into account the transverse part. 
The corresponding scalar parts of the selfenergy can be extracted 
by applying the transverse projection operator from eq.~(\ref{PLT}): 
\begin{eqnarray} 
\Sigma_{\rho\alpha}^{(0)}(q_0,q) & \equiv & \frac{1}{2} \  
(P_{T})_{\mu\nu} \ \Sigma_{\rho\alpha}^{(0),ij}(q_0,\vec q) 
\nonumber\\
 & = &  \frac{1}{2} \left(\frac{f_{\rho\alpha} \ 
F_{\rho\alpha}(q)}{m_\rho}
\right)^2 \ SI(\rho\alpha) \ q^2 \ \phi_{\rho\alpha}(q_0,q)
\nonumber\\
 & \equiv & -q^2 \ \chi_{\rho\alpha}^{(0)}(q_0,q) \ , 
\end{eqnarray} 
where the spin-isospin factors are summarized in table~\ref{tab2}. 
Guided by the experience one has from pion nuclear physics, we 
also account for short-range correlation effects in the $\rho$-like 
particle-hole bubbles, parametrized by Migdal parameters $g'$. The
renormalized  selfenergy in a given excitation channel then becomes
\begin{equation}
\Sigma_{\rho\alpha}(q_0,q)=-q^2 \ 
\frac{\chi_{\rho\alpha}^{(0)}(q_0,q)}
{1+g'_\alpha \ \chi_{\rho\alpha}^{(0)}(q_0,q)} \ . 
\end{equation} 
For more than one channel one again has to solve a matrix
problem as given by eq.~({\ref{chi}) 
\begin{equation} 
\chi_{\rho\alpha}  =  \chi_{\rho\alpha}^{(0)}-\sum_{\beta} \
\chi_{\rho\alpha}^{(0)} \ {g'}_{\alpha\beta} \ \chi_{\rho\beta} \ .
\label{chirho}
\end{equation}
Both the $N(1720)$ and $\Delta(1905)$ resonances have sufficiently 
high masses which allow to disentangle their coupling strength to 
the $\rho N$ channel through the corresponding decay process.  However, 
within our full off-shell treatment of the in-medium $\rho$ propagator 
we are also able to incorporate lower-lying $\rho$-like particle-hole
excitations. The most obvious candidate is, of course, the 
$\Delta(1232)$: it's typical excitation energy of about 300~MeV closely 
coincides with the invariant mass region where the experimentally 
observed excess of dilepton radiation in heavy-ion collisions 
at CERN energies sets in~\cite{CERES,HELIOS3}. 
An appropriate Lagrangian is given by  
\begin{equation} 
{\cal L}_{\rho\Delta N}  =  \frac{f_{\rho \Delta N}}{m_\rho} \
\Psi^\dagger_{\Delta} \ ({\vec S}_\frac{3}{2} \times {\vec q}) \cdot
({\bf T} \cdot \vec{\pmb{$\rho$}}) \ \Psi_N \  + \ {\rm} h.c. \ , 
\end{equation} 
which automatically includes $\rho N\Delta^{-1}$ excitations as 
well. To complete the picture of low-lying $\rho$-baryon excitations 
we furthermore account for $NN^{-1}$ and $\Delta\Delta^{-1}$ bubbles 
described by 
\begin{eqnarray} 
{\cal L}_{\rho NN} & = & \frac{f_{\rho NN}}{m_\rho} \ \Psi^\dagger_N \ 
({\vec \sigma} \times {\vec q}) \cdot (\pmb{$\tau$} \cdot 
\vec{\pmb{$\rho$}}) \ \Psi_N
\nonumber\\ 
{\cal L}_{\rho \Delta\Delta} & = & \frac{f_{\rho\Delta\Delta}}{m_\rho} 
 \ \Psi^\dagger_\Delta \ 
(2{\vec \Sigma}_{\frac{3}{2}} \times {\vec q}) \cdot 
(\pmb{$\Theta$} \cdot \vec{\pmb{$\rho$}}) \ \Psi_\Delta \ ,  
\label{Lrnndd} 
\end{eqnarray} 
where ${\vec \Sigma}_{\frac{3}{2}}$ and $\pmb{$\Theta$}$ denote the 
spin-$\frac{3}{2}$ and isospin-$\frac{3}{2}$ operators, respectively.  
The coupling constant $f_{\rho NN}$ is taken from the Bonn potential, 
and $f_{\rho N\Delta}$=2$f_{\rho NN}$, 
$f_{\rho \Delta\Delta}$=$\frac{1}{5}f_{\rho NN}$, 
cp. table~\ref{tab2}. Including Migdal 
parameters as in the pionic case (all additional ones are uniformly 
set to $g'_{\alpha\beta}$=0.5), the total selfenergy for $\rho$-baryon 
interactions in the particle-hole picture becomes 
\begin{equation} 
\Sigma_{\rho BB^{-1}}(q_0,q)= -q^2 \ \sum_\alpha \chi_{\rho\alpha}(q_0,q)  
\label{sgrbb} 
\end{equation}   
where the summation is over $\alpha=NN^{-1},\Delta N^{-1},
N\Delta^{-1},\Delta\Delta^{-1},N(1720)N^{-1},\Delta(1905)N^{-1}$, and 
the full susceptibilities are determined from solving 
eq.~(\ref{chirho}).   

In fig.~5 we display the transverse part of the $\rho$ spectral 
function, 
\begin{equation} 
{\rm Im} D_\rho^T(M,q)= {\rm Im} \left( \frac{1}{M^2-(m_\rho^{bare})^2 
-\Sigma_{\rho\pi\pi}^0(M)-\Sigma_{\rho BB^{-1}}(q_0,q)} \right) \ ,  
\end{equation} 
at fixed baryon chemical potential and with no medium effects included 
in the 2-pion states. With both increasing density (upper panel in 
fig.~5) and 3-momentum (lower panel in fig.~5) the 'rhosobar' 
branches build up an appreciable enhancement for invariant masses 
below $M$$\simeq$0.6~GeV, together with a depletion of the $\rho$ peak.
However, due to the thermal motion of the nucleons and the 
relatively large widths of the $N(1720)$, $\Delta(1905)$ and 
$\Delta(1232)$ resonances, the overlapping branch structures are
hardly visible; only at the highest nucleon density considered 
($\rho_N$=0.7$\rho_0$ at $T$=0.17~GeV, dotted line in the upper 
panel of fig.~5) one recognizes a shoulder around $M$=0.6~GeV 
corresponding to the $\rho N^*N$-channel. 

%%%%%%%%%%%%%%%%%%%%%%%%%%%%%%%%%%%%%%%%%%%%%%%%%%%%%%%%%%%%%%%%%%%%%%%%
\section{Rho Scattering in a Thermal Meson Gas}  
%%%%%%%%%%%%%%%%%%%%%%%%%%%%%%%%%%%%%%%%%%%%%%%%%%%%%%%%%%%%%%%%%%%%%%%%
In central collisions of heavy nuclei at CERN energies (160-200~GeV/u)
large numbers of secondary mesons are produced, clearly exceeding 
the number of primary nucleons. However, as one knows from the pion, 
it's properties are typically much less affected in a hot meson gas 
than in the nuclear medium at comparable densities~\cite{RPhD}. 
Nevertheless, a reliable description of $\rho$ meson properties 
at CERN-SpS energies certainly has to account for interactions
with surrounding mesons. An  analysis of collision 
rates for (on-shell) $\rho$-mesons in a $\pi$/$K$ gas has been 
performed in ref.~\cite{Hagl}, where a moderate collisional 
broadening of $\bar\Gamma_\rho^{coll}$($T$$\le$0.17~GeV)$\le$60~MeV 
has been found. \\
Our point here is to include $\rho\pi$ and $\rho K$ scattering as an 
additional contribution to the (off-shell) $\rho$ selfenergy in 
hot hadronic matter. In analogy to the previous section we assume the 
dominant contribution to arise from s-channel resonance formation, 
{\it i.e.} $\rho\pi\to a_1(1260)$ and $\rho K/\rho\bar K \to K_1(1270)/
\bar K_1(1270)$. For the corresponding interaction Lagrangian we 
choose the one proposed in ref.~\cite{XSB}, which is (isospin structure
suppressed) 
\begin{equation} 
{\cal L}_{\pi\rho a_1}=G_{\pi\rho a_1} \ a_\mu \ (g^{\mu\nu} \ q\cdot 
p_\pi - q^\mu p_\pi^\nu) \ \rho_\nu \ \pi \ , 
\label{Lpirhoa1} 
\end{equation}  
being compatible with gauge invariance within the VDM approach. 
$a_\mu$, $\rho_\nu$ and $\pi$ denote the fields for the $a_1(1260)$
meson, rho meson and pion, and $q$, $p_\pi$ the rho and pion 4-momenta, 
respectively.  
Similarly, we describe the $K\rho K_1(1270)$ interaction by 
\begin{equation} 
{\cal L}_{K\rho K_1}=G_{K\rho K_1} \ b_\mu \ (g^{\mu\nu} \ q\cdot
p_K - q^\mu p_K^\nu) \ \rho_\nu \ K 
\label{LkarhoK1} 
\end{equation}  
($K$:~kaon field, $b_\mu$:~$K_1(1270)$ field). 
Along the same lines as in the previous section we fix the coupling 
constants $G$ by adjusting the partial decay widths 
$\Gamma_{a_1\to\pi\rho}$, $\Gamma_{K_1\to K \rho}$ to their experimental
values. Including the finite width of the $\rho$ meson one has 
\begin{eqnarray} 
\Gamma_{a_1\to\pi\rho}(\sqrt{s}) & = & \frac{G_{\pi\rho a_1}^2}{24\pi s}
\int\limits_{2m_\pi}^{M_{max}} \frac{M dM}{\pi} \  A_\rho(M)
\nonumber\\ 
 & & \qquad\qquad\quad * \ q_{cm} \ [\frac{1}{2}(s-M^2-m_\pi^2)^2
+M^2(m_\pi^2+q_{cm}^2)] \ F_{\pi\rho a_1}(q_{cm})^2 
\label{gammaa1}
\end{eqnarray}
with a hadronic dipole form factor $F_{\pi\rho a_1}(q_{cm})$ 
($\Lambda_{\pi\rho a_1}$=2~GeV) and $M_{max}=\sqrt s-m_\pi$. 
Assuming $\Gamma_{a_1\to\pi\rho}(\sqrt{s}=m_{a_1})\equiv 
\Gamma_{a_1}^{tot}$=0.4~GeV with $m_{a_1}$=1.23~GeV yields  
$G_{\pi\rho a_1}$=18.7~GeV$^{-1}$, which is not very different from 
the value of 14.8~GeV$^{-1}$ obtained for a zero-width $\rho$ meson
in ref.~\cite{XSB}. An analogous calculation for the $K_1$ width, 
taking $\Gamma_{K_1\to K\rho}(\sqrt{s}=m_{K_1})$=0.06~GeV and 
$m_{K_1}$=1.27~GeV~\cite{PDG}, gives $G_{K\rho K_1}$=15.3~GeV$^{-1}$. 

Within the imaginary time (Matsubara) approach the $\rho$ meson 
selfenergy tensor arising from interactions with surrounding pions 
at temperature $T$ can be calculated as 
\begin{equation} 
\Sigma_{\rho\pi}^{\mu\nu}(q_0,\vec q)=\int \frac{d^3p}{(2\pi)^3} 
\frac{1}{2\omega_p^\pi} [f^\pi(\omega_p^\pi)-
f^{\pi\rho}(\omega_p^\pi+q_0)] \ M_{\pi\rho}^{\mu\nu}(p_\pi,q) 
\end{equation}  
with bosefactors $f^\pi$, $f^{\pi\rho}$ and the isospin averaged 
forward scattering amplitude $M_{\pi\rho}^{\mu\nu}$. Assuming the
latter to be dominated by $a_1$ resonance formation one
derives from eq.~(\ref{Lpirhoa1}):
\begin{eqnarray} 
M_{\pi\rho a_1}^{\mu\nu}(p_\pi,q) & = & 
\frac{(2I_{a_1}+1)}{(2I_\rho+1)} \ 
G_{\pi\rho a_1}^2 \ F_{\pi\rho a_1}(q_{cm})^2 
\nonumber\\ 
 & &  \qquad \qquad 
* (g^{\mu\kappa} \  p_\pi \cdot q- p_\pi^\mu q^\kappa)
 \ D_{a_1,\kappa\lambda}(k) \ (g^{\lambda\nu} q \cdot p_\pi - 
q^\lambda p_\pi^\nu) \ , 
\label{Mpirho} 
\end{eqnarray} 
where the $a_1$ propagator at 4-momentum $k\equiv(p_\pi+q)$ is given by
\begin{equation}
D_{a_1,\kappa\lambda}(k)=
\frac{(g_{\kappa\lambda}-k_\kappa k_\lambda / m_{a_1}^2)}
{s-m_{a_1}^2+im_{a_1}\Gamma_{a_1}^{tot}(s)}  
\end{equation} 
($s=k_\lambda k^\lambda$). The total $a_1$ width $\Gamma_{a_1}^{tot}(s)$
is taken from eq.~(\ref{gammaa1}). Using the projection operators 
defined in eq.~(\ref{PLT}) we can extract the explicit form of the 
longitudinal and transverse part of the selfenergy as 
\begin{eqnarray} 
\Sigma_{\rho\pi a_1}^{L}(q_0,q) & = & (P_L)_{\mu\nu} \ 
\Sigma_{\rho\pi a_1}^{\mu\nu}(q_0,\vec q) 
\nonumber\\
 & = & G_{\pi\rho a_1}^2 \ \int\frac{p^2 dp dx}{(2\pi)^2 2\omega_p^\pi}
 \ [f^\pi(\omega_p^\pi)-f^{\pi\rho}(\omega_p^\pi+q_0)] \ 
 F_{\pi\rho a_1}(q_{cm})^2
\nonumber\\ 
 & & \quad  * \frac{[\frac{1}{4}(s-M^2-m_\pi^2)^2 
\frac{M^2}{m_{a_1}^2}+ {\omega_p^\pi}^2 M^2 (1-\frac{M^2}{m_{a_1}^2})
-\vec p^2 M^2 (1-\frac{M^2}{m_{a_1}^2}) x^2]}
{s-m_{a_1}^2+im_{a_1} \Gamma_{a_1}^{tot}}
\nonumber\\
\Sigma_{\rho\pi a_1}^{T}(q_0,q) & = & \frac{1}{2} \ (P_T)_{\mu\nu} \
\Sigma_{\rho\pi a_1}^{\mu\nu}(q_0,\vec q)
\nonumber\\
 & = & \frac{1}{2}  \ G_{\pi\rho a_1}^2  \ 
\int\frac{p^2 dp dx}{(2\pi)^2 2\omega_p^\pi}
\ [f^\pi(\omega_p^\pi)-f^{\pi\rho}(\omega_p^\pi+q_0)]
\  F_{\pi\rho a_1}(q_{cm})^2
\nonumber\\
 & & \quad * \frac{[\frac{1}{2}(s-M^2-m_\pi^2)^2
\frac{M^2}{m_{a_1}^2}-\vec p^2 M^2 (1-\frac{M^2}{m_{a_1}^2}) (1-x^2)]}
{s-m_{a_1}^2+im_{a_1} \Gamma_{a_1}^{tot}}
\label{sgrpa} 
\end{eqnarray}   
with $x=\cos\theta$, $\theta=\angle(\vec p, \vec q)$. Completely 
analogous expressions are obtained for $K\rho$ scattering 
via $K_1(1270)$ resonance formation (replacing 
$\pi\to K$, $a_1\to K_1$; note that the isospin averaging factor in 
eq.~(\ref{Mpirho}) changes from 1 to 2/3 and that the total width 
entering $D_{K_1}$ is taken to be 
$\Gamma_{K_1}^{tot}$=2$\Gamma_{K_1\to K\rho}$; for antikaons one has 
$\Sigma_{\rho \bar K \bar K_1}$=$\Sigma_{\rho KK_1}$ as long as 
$\mu_{\bar K}$=$\mu_{K}$). 
The spin-averaged $\rho$ propagator in a hot $\pi$-$K$ gas is then 
given by 
\begin{eqnarray} 
D_\rho(M,q) & = & \frac{1}{3} \ [(P_L)_{\mu\nu}+(P_T)_{\mu\nu}] \ 
D_\rho^{\mu\nu}(M,\vec q)
\nonumber\\ 
& = & \frac{1}{3} \ [-g_{\mu\nu}+\frac{q_\mu q_\nu}{M^2} ] \ 
D_\rho^{\mu\nu}(M,\vec q)
\nonumber\\
 & = & \frac{1}{3} \left( \frac{1}{M^2-(m_\rho^{bare})^2-
\Sigma_{\rho\pi\pi}^0(M)-\Sigma_{\rho\pi a_1}^L(M,q)-
2\Sigma_{\rho KK_1}^L(M,q)} \right. 
\nonumber\\
 & & \left. \quad + \frac{2}{M^2-(m_\rho^{bare})^2-
\Sigma_{\rho\pi\pi}^0(M)-\Sigma_{\rho\pi a_1}^T(M,q)-
2\Sigma_{\rho KK_1}^T(M,q)}  \right) \ , 
\label{drhopik}  
\end{eqnarray}  
the imaginary part of which is shown in fig.~6 for $\mu_\pi=\mu_K=0$. 
With increasing temperature (upper panel) $\pi\rho$ and 
$K\rho$/$\bar K\rho$ scattering simply
lead to a moderate broadening of the spectral function, which is 
similar to the results of ref.~\cite{Hagl}. The impact of a finite
3-momentum is rather small (lower panel of fig.~6).   

%%%%%%%%%%%%%%%%%%%%%%%%%%%%%%%%%%%%%%%%%%%%%%%%%%%%%%%%%%%%%%%%%%%%%%%%
\section{\lowercase{$e^+e^-$} Spectra in Central Heavy-Ion Collisions 
at CERN/SpS Energies}
%%%%%%%%%%%%%%%%%%%%%%%%%%%%%%%%%%%%%%%%%%%%%%%%%%%%%%%%%%%%%%%%%%%%%%%%
The general expression for the dilepton production rate 
R=$dN_{l^+l^-}/d^4x$ per unit of 4-momentum $d^4q$ in a hadronic 
medium of temperature $T$ can be written as
\begin{equation} 
\frac{dN_{l^+l^-}}{d^4xd^4q}=L_{\mu\nu}(q) H^{\mu\nu}(q) \ .  
\end{equation} 
To lowest order in the electromagnetic coupling $\alpha$=1/137 the 
lepton tensor is given by 
\begin{equation} 
L_{\mu\nu}(q)=-\frac{\alpha^2}{3\pi^2 M^2} \left( g_{\mu\nu} -
\frac{q_\mu q_\nu}{M^2} \right) \ , 
\end{equation} 
where we neglected the rest mass of the leptons compared to their 
individual 3-momenta $|\vec p_+|$, $|\vec p_-|$ (in the following we 
will focus on the $e^+e^-$ case). $q=p_++p_-$ is the total 4-momentum 
of the pair in the heat bath.  \\
To calculate the dilepton prodcution rate from the 
$\pi^+\pi^-\to e^+e^-$ annihilation process we employ the 
phenomenologically well established vector dominance model (VDM). 
It relates the hadronic part of the electromagnetic current to the
third component of the isovector $\rho$ meson field as 
\begin{equation}
J^\mu=\frac{(m_\rho^{bare})^2}{g} \rho_3^\mu \ , 
\end{equation} 
with the universal VDM coupling constant $g=g_{\rho\pi\pi}$. As a 
consequence, the hadronic tensor $H^{\mu\nu}$ can be expressed in terms
of the imaginary part of the retarded $\rho$ propagator in hot/dense 
matter: 
\begin{equation}
H^{\mu\nu}(q_0, \vec q;\mu_B,T)=-f^\rho(q_0;T) \ \frac{(m_\rho^{bare})^4}
{\pi g_{\rho\pi\pi}^2} \ {\rm Im}D_\rho^{\mu\nu}(q_0,\vec q;\mu_B,T) \ 
\end{equation}
($f^\rho(q_0,T)=(e^{q_0/T}-1)^{-1}$).
On account of current conservation, $q_\mu {\rm Im} D_\rho^{\mu\nu}=0$,
and using the decomposition eq.~(\ref{drhomunu}) we arrive at 
\begin{equation}
{dN_{\pi^+\pi^-\to e^+e^-}\over d^4xd^4q} =
-\frac{\alpha^2 (m_\rho^{bare})^4}{\pi^3 g_{\rho\pi\pi}^2} \ 
\frac{f^\rho(q_0;T)}{M^2} \ \frac{1}{3} \ 
({\rm Im}D_\rho^L(q_0,q;\mu_B,T)+2 \ {\rm Im}D_\rho^T(q_0,q;\mu_B,T))  
\label{rate}
\end{equation}
with 
\begin{equation} 
{\rm Im}D_\rho^{L,T}(q_0,q;\mu_B,T)=
\frac{{\rm Im} \Sigma_\rho^{L,T}(q_0,q;\mu_B,T)}{|M^2-(m_\rho^{bare})^2 
-\Sigma_\rho^{L,T}(q_0,q;\mu_B,T)|^2}  \ . 
\end{equation} 
The full $\rho$ meson selfenergy is obtained from combining the 
results of the three preceeding sections, eqs.~(\ref{imsgrpp}), 
(\ref{resgrpp}), (\ref{sgrbb}) and (\ref{sgrpa}): 
\begin{eqnarray} 
\Sigma_\rho^L & = & \Sigma_{\rho\pi\pi}+\Sigma_{\rho\pi a_1}^L+
\Sigma_{\rho KK_1}^L + \Sigma_{\rho \bar{K}\bar{K}_1}^L 
\nonumber\\
\Sigma_\rho^T & = & \Sigma_{\rho\pi\pi}+ \Sigma_{\rho BB^{-1}} 
+\Sigma_{\rho\pi a_1}^T+ \Sigma_{\rho KK_1}^T
+\Sigma_{\rho \bar{K}\bar{K}_1}^T \ .  
\end{eqnarray} 
In fig.~7 we show our final result for the spin averaged $\rho$ spectral
function Im~$D_\rho\equiv \frac{1}{3} ({\rm Im} D_\rho^L + 
2{\rm Im} D_\rho^T)$, which does not exhibit any principly new 
aspects as compared to the analyses in the previous sections. 

\noindent 
For calculating $e^+e^-$ invariant mass spectra as measured in the 
CERES experiment the  differential rate eq.~(\ref{rate}) has to be 
integrated over 3-momentum as well as the space-time history of a 
central 200~GeV/u S-Au reaction. Therefore we make use of recent 
transport calculations~\cite{LKB} according to which the corresponding 
temperature evolution can be well represented by an exponential 
decrease 
\begin{equation} 
T(t)=(T^i-T^\infty) \ e^{-t/\tau} + T^\infty
\label{temp} 
\end{equation}
with an initial temperature $T^i$=0.170~GeV, $\tau$=8~fm/c and 
$T^\infty$=0.110~GeV. The initial baryon density is predicted 
by RQMD simulations to be $\rho_B^i$$\simeq$2.5$\rho_0$~\cite{Sorge}.
When including the 12 lowest lying baryonic resonances this corresponds
to a baryon chemical potential of $\mu_B$=0.39~GeV, which is slightly  
lower than the value we employed in~\cite{CRW}. If we furthermore 
assume chemical equilibrium, {\it i.e.} $\mu_B$=const and  
$\mu_{meson}$$\equiv$0, the time evolution of baryon and meson 
densities is completely determined by eq.~(\ref{temp}). The such 
obtained baryon density $\rho_B(t)$, {\it e.g.}, turns out to be 
in good agreement with the transport simulations.   
After a variable transformation $q_0 dq_0 = M dM$ and with an 
isotropic density profile of the hadronic fireball at each stage 
of the expansion, eq.~(\ref{rate}) becomes   
\begin{eqnarray}  
\frac{dN_{\pi^+\pi^-\to e^+e^-}}{dM d\eta} & = &  
N_0 \ \frac{\alpha^2 (m_\rho^{bare})^4}{\pi^3 g_{\pi\pi\rho}^2 M^2} \ 
\int\limits_0^{t_{fo}} dt \ V(t)
\nonumber\\  
 & & \qquad \qquad *  \int d^3q \ \frac{M}{q_0} \ f^\rho(q_0;T(t)) \ 
{\rm Im} D_\rho(M,q; \mu_B,T(t)) \ A(M,\vec q) \ .   
\label{spec} 
\end{eqnarray}
$A(M,\vec q)$ accounts for the experimental acceptance cuts on the 
dilepton tracks as applied in the CERES detector ({\it i.e.} 
$p_T$$>$0.2~GeV, 2.1$<$$\eta$$<$2.65 and $\Theta_{e^+e^-}$$>$35~mrad).
We here also include it's finite mass resolution by folding our 
theoretically calculated spectrum with a gaussian 
\begin{equation}
G(M,m)=\frac{1}{\sigma(M) \sqrt{2\pi} } 
\exp \left( -\frac{(m-M)^2} {2\sigma(M)^2} \right) \ ,  
\end{equation} 
where the width $\sigma(M)$ is extracted from ref.~\cite{DrUl}.   
The dimensionless normalization constant $N_0$ is fixed using free 
$\rho$ decays by matching our corresponding total ($M$-integrated) 
$e^+e^-$-yield with the transport results of Li, Ko and 
Brown~\cite{LKB}. In eq.~(\ref{spec}) $t_{fo}$ denotes the freezeout
time of the hadronic fireball. For central S-Au reactions we use 
$t_{fo}$=10~fm/c (corresponding to a baryonic freezeout density 
$\rho_B(t_{fo})$=0.32$\rho_0$), but our results are rather insensitive
with respect to moderate variations in $t_{fo}$. The fireball volume
$V(t)$ is determined by the baryon density as 
\begin{equation} 
V(t)=\frac{N_B}{\rho_B(t)} \ ,
\end{equation} 
where $N_B$ denotes the number of participating baryons (which can be 
absorbed in the normalization constant $N_0$).   \\
In fig.~8 we compare the resulting dielectron spectrum, 
supplemented with contributions 
from free Dalitz decays ($\pi_0$,$\eta$~$\to$~$\gamma e^+e^-$, 
$\omega$~$\to$~$\pi^0e^+e^-$) and free $\omega$~$\to$~$e^+e^-$ decays
as extracted from ref.~\cite{LKB}, with the CERES data from central 
S-Au collisions at 200~GeV/u. Apparently, the 
combined medium effects in the $\rho$ propagation lead to quite 
reasonable agreement with the experimental spectrum. As was identified
in the preceeding sections, both the dressing of the intermediate
2-pion states as well as particle-hole-type excitations by the 
$\rho$ meson ('rhosobars') constitute the major part of the effect.

We also performed a similar analysis for central Pb-Au collisions 
at 158~GeV/u. Here, RQMD predicts initial temperatures and baryon 
densities of $T^i$$\simeq$0.180~GeV and 
$\rho_B^i$$\simeq$4$\rho_0$~\cite{Sorge}, which corresponds to a 
baryon chemical potential of $\mu_B$=0.408~GeV.  Taking $\tau$=10~fm/c
and $T^\infty$=0.105~GeV in eq.~(\ref{temp}), the time evolution 
of temperature and 
density as found in transport calculations~\cite{LKBS} can again be 
well reproduced. The much larger system size as compared to the 
S-Au case increases the fireball lifetime substantially. We account 
for this by choosing a freezeout time of $t_{fo}$=20~fm/c 
corresponding to a freezeout temperature and density of 
$T^{fo}$=0.115~GeV and $\rho_B(t_{fo})$=0.18$\rho_0$, respectively.   
This is somewhat lower than in S-Au collisions, which reflects the fact 
that the hadronic cooling becomes more efficient with increasing  
system size. The comparison of the resulting dielectron spectrum 
with preliminary data from the CERES/NA45 collaboration confirms 
our findings for S-Au case: hadronic rescattering in  
in-medium $\rho$ propagation seems to resolve the discrepancy 
between the experimental data and theoretical results based on  
the free $\pi^+\pi^-$ annihilation process.   
 
%%%%%%%%%%%%%%%%%%%%%%%%%%%%%%%%%%%%%%%%%%%%%%%%%%%%%%%%%%%%%%%%%%%%%%%%
\section{Summary and Conclusions}
%%%%%%%%%%%%%%%%%%%%%%%%%%%%%%%%%%%%%%%%%%%%%%%%%%%%%%%%%%%%%%%%%%%%%%%%
Starting from a realistic model for the $\rho$ meson in free space we 
have studied $\rho$ propagation in hot hadronic matter and it's impact 
on dilepton production in URHIC's at CERN-SpS energies. Since the 
free $\rho$ strongly couples to $\pi\pi$ states, an important medium 
effect arises from the modification of the pion propagation: a strong 
softening of the pion dispersion relation, generated by the coupling to 
$\Delta N^{-1}$ as well as low-lying $NN^{-1}$, $\Delta\Delta^{-1}$ 
excitations and further weighted by thermal occupation factors in 
the hot gas, leads to an appreciable enhancement of the $\rho$ spectral 
function from zero to about 0.6~GeV in invariant mass.   
% The pionic collectivity in the RPA-type summation of the 
% particle-hole bubbles turned out to be essential for this effect. 
However, as we have shown earlier~\cite{CRW}, these modifications do 
not yet allow for a quantitative description of the CERES data. 
We furthermore included {\it direct} interactions of the $\rho$ within 
the hot hadronic enviroment. Corresponding scattering processes 
off surrounding nucleons, deltas, pions and kaons, were assumed  
to be dominated by s-channel resonance graphs, which we derived 
in accordance with constraints from gauge invariance. Within a full 
off-shell treatment the most notable contributions come from 
'rhosobar'-type $\Delta N^{-1}$, $N(1720)N^{-1}$ and 
$\Delta(1905)N^{-1}$ excitations, which result in a further shift 
of strength to low invariant masses in the $\rho$ spectral function.  
$a_1(1260)$ and $K_1(1270)$ resonance formation in $\rho\pi$ and 
$\rho K$ scattering turned out to be less significant.
Within the VDM we applied our full $\rho$ propagator to compute 
$e^+e^-$ production rates from in-medium $\pi\pi$ annihilation. These  
were integrated according to a temperature/density evolution 
of central S-Au (200~GeV/u) and Pb-Au (158~GeV/u) collisions as found 
in recent transport calculations, including experimental acceptance 
conditions.  It turns out that the observed enhancement in the 
CERES data can be accounted for. \\ 
One can certainly think of other medium effects in $\rho$ propagation
that are not included in our present study. Among those are, 
{\it e.g.}, finite temperature modifications of the VDM ({\it i.e.} 
corrections to the $\rho\gamma$ vertex), or $\sigma$-tadpole 
graphs~\cite{FrSo}, which represent the coupling of the $\rho$ meson 
to a scalar field generated by the surrounding hadrons. Furthermore, 
the vector dominance assumption for the nucleonic sector is 
presumably less accurate than for the pion, which might overestimate 
the effects of the baryonic medium somewhat (especially for 
very low invariant masses).   However, we do not expect these 
shortcomings to affect our results in a major, {\it qualitative} way. 
Thus, it seems justified to conclude that medium modifications of the  
$\rho$ meson arising from hadronic rescattering can essentially explain 
the low-mass dilepton enhancement as found in recent experiments at the 
CERN-SpS. After all, the Brown-Rho conjecture of 
a dropping $\rho$ mass might in fact be a related phenomenon, 
{\it i.e.} in some sense an efficient way to parametrize the shift 
of strength to low invariant masses in the $\rho$ spectral function. 
However, in our dynamical calculation this shift of strength is 
due to a strong broadening of the resonance, whereas
BR-Scaling predicts a downwards shift of the peak caused by a strong 
scalar mean field in the hadronic medium. Some  
of this discrepancy might be resolved when including the $\sigma$ 
tadpole graph in our framework. \\ 
Further refinements also require 
to include a finite $\vec q$-dependence of the 2-pion selfenergy. 
Finally, we need to improve our treatment of the URHIC-dynamics; here,
a hydrodynamic approach, which accounts for the full off-shell behavior 
implicit in our dilepton production rate, is mandatory.   
Work in all these directions is in progress.

\vskip1cm
 
\centerline {\bf ACKNOWLEDGMENTS}
We are grateful for productive conversations with  G.E. Brown, 
A. Drees, J.W. Durso, B. Friman, C.M. Ko, E.V. Shuryak and H. Sorge. 
We are indebted to G.Q. Li for providing us with the transport
results for free meson decays. 
One of us (RR) acknowledges financial support from the 
Alexander-von-Humboldt foundation as a Feodor-Lynen fellow. 
This work is supported in part by the National Science Foundation
under Grant No. NSF PHY94-21309 and by the U.S. Department of Energy 
under Grant No. DE-FG02-88ER40388.

\newpage

\begin{table}
\caption{\em Spin-isospin transition factors and coupling constants for
pion induced (longitudinal) p-wave particle-hole excitations in a hot 
$N\Delta$ gas.}
\label{tab1}
\begin{tabular}{ccccc}
 $\pi\alpha$ & $\pi NN^{-1}$ & $\pi\Delta N^{-1}$ & 
$\pi N\Delta^{-1}$ & $\pi\Delta\Delta^{-1}$ \\
\hline
$SI(\pi\alpha)$ & 4 & 16/9 & 16/9 & 400 \\
$f_{\pi\alpha}^2/4\pi$ & 0.081 & 0.324 & 0.324 & 0.00324 \\
\end{tabular}
\end{table}
\begin{table}
\caption{\em Spin-isospin transition factors and coupling constants for
$\rho$ meson induced (transverse) p-wave particle-hole excitations in 
a hot $N\Delta$ gas.}
\label{tab2}
\begin{tabular}{ccccccc}
 $\rho\alpha$ & $\rho N(1720)N^{-1}$  & $\rho \Delta(1905)N^{-1}$ & 
 $\rho\Delta N^{-1}$ & $\rho N\Delta^{-1}$ & $\rho NN^{-1}$ & 
 $\rho\Delta\Delta^{-1}$ \\
\hline
$SI(\rho\alpha)$ & 16/3 & 8/5 & 32/9 & 32/9 & 8 & 800 \\
$f_{\rho\alpha}^2/4\pi$ & 6.99 & 10.64 & 18.72  & 18.72 & 4.68 & 0.19 \\
\end{tabular}
\end{table}

\pagebreak

%
%
%%%%% Figure Captions 
\begin{center}
{\large \sl \bf Figure Captions}
\end{center}
\vspace{0.5cm}

\begin{itemize}
\item[{\bf Figure 1}:]  
Our fit to the p-wave $\pi\pi$ scattering phase shifts (upper panel) 
and the pion electromagnetic form factor in free space (lower panel);  
the squares in the lower panel correspond to the Gounaris-Sakurai 
formula~\cite{GoSa}, which gives an accurate description of the data. 

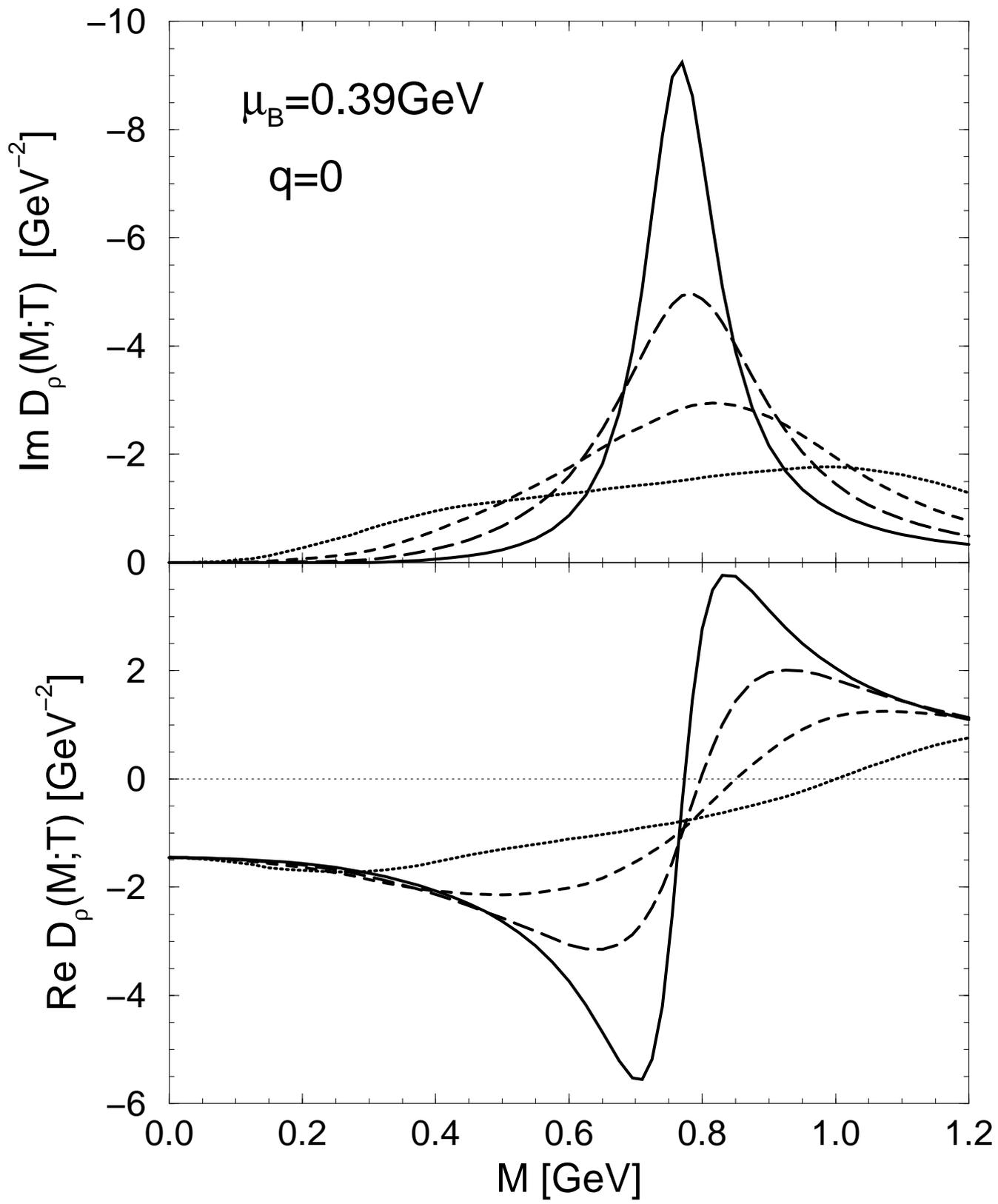
\item[{\bf Figure 2:}]  
Impact of in-medium $\pi\pi$ states on the $\rho$ propagator 
(upper panel: imaginary part, lower panel: real part) at zero 
3-momentum in a hot $\pi N\Delta$ with chemical potentials 
$\mu_B$$\equiv$$\mu_N$=$\mu_\Delta$=0.39~GeV and $\mu_\pi$=0. The 
chosen temperatures 
of $T$=0.170~GeV (dotted lines), $T$=0.149~GeV (short-dashed lines) and 
$T$=0.127~GeV (long-dashed lines) correspond to $N\Delta$ densities of 
$\rho_{N\Delta}/\rho_0$=1.41, 0.63 and 0.32, respectively. The solid 
lines represent the results in free space. 

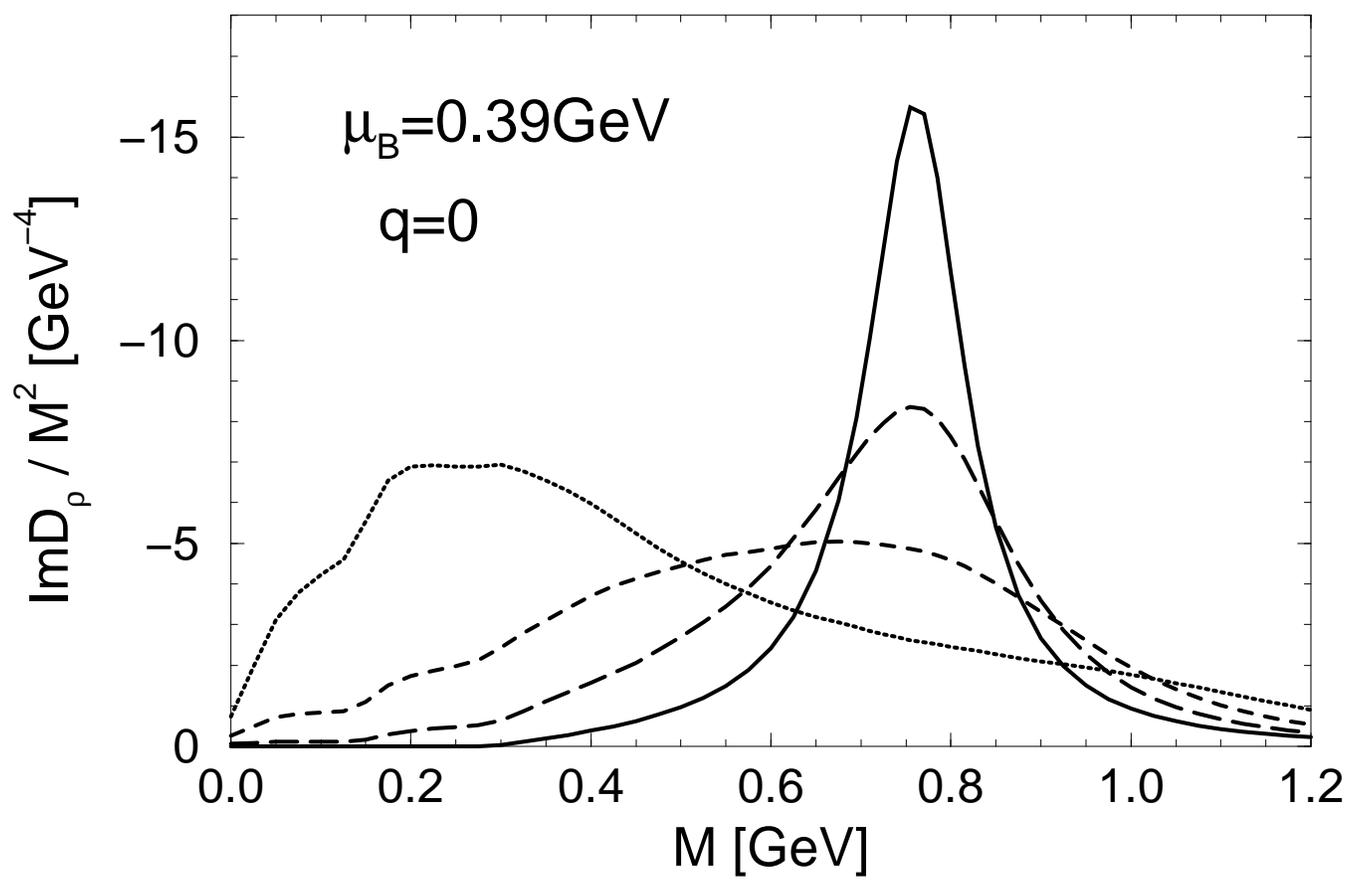
\item[{\bf Figure 3:}] 
Imaginary part of the $\rho$ propagator, weighted with a photon 
propagator 1/$M^2$, in a hot $\pi N\Delta$ gas; medium modifications 
are due to the dressing of the intermediate 2-pion states; line 
identification as in fig.~2. 

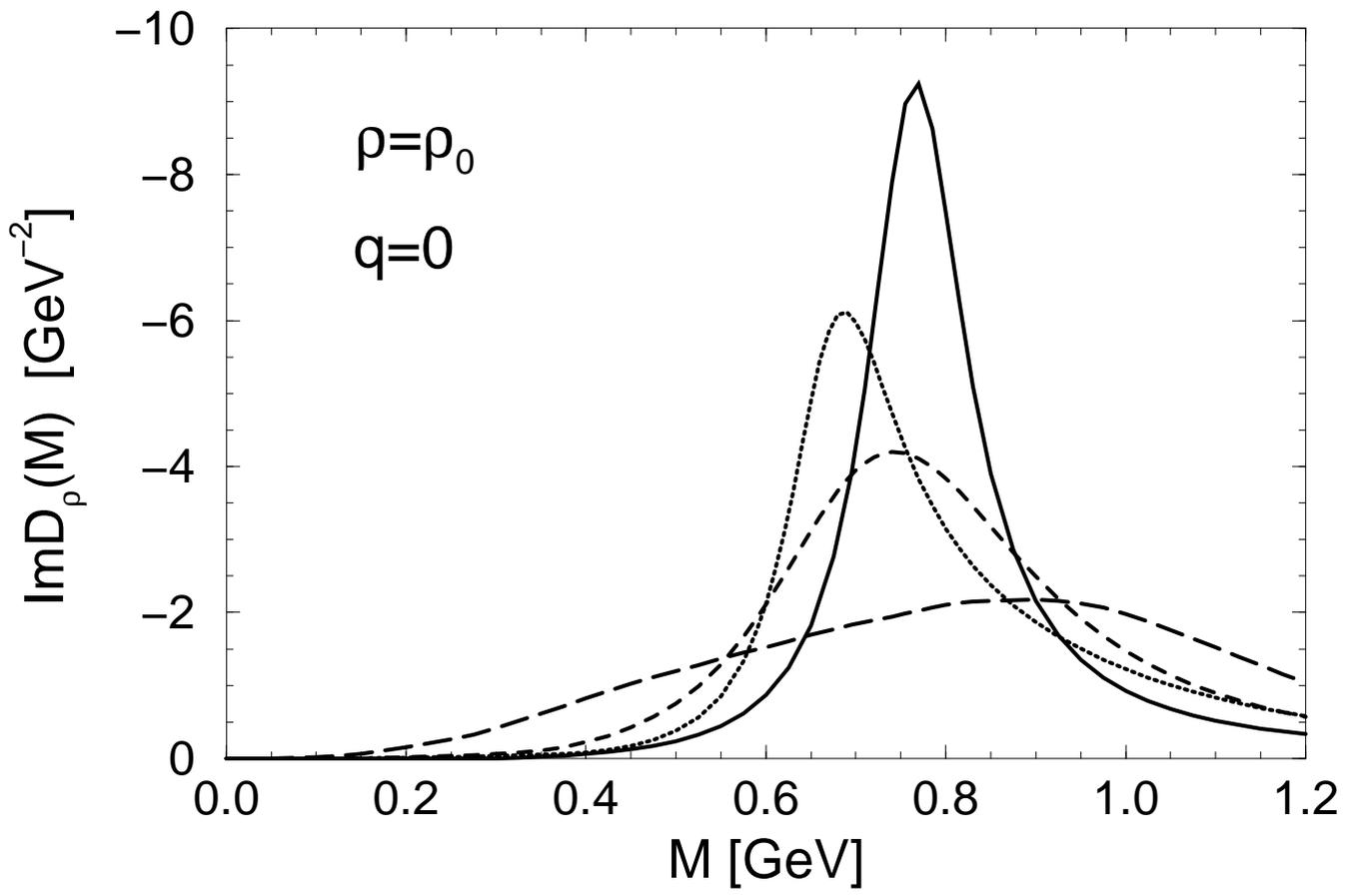
\item[{\bf Figure 4:}]
Imaginary part of the $\rho$ propagator with in-medium $\pi\pi$ states
at normal nuclear matter density; \\
long-dashed line: full result at $T$=0.150~GeV; short-dashed line: 
without pionic collectivity (as described in the text) at $T$=0.150~GeV;
dotted line: without pionic collectivity at $T$=0.050~GeV; full line: 
free space.  

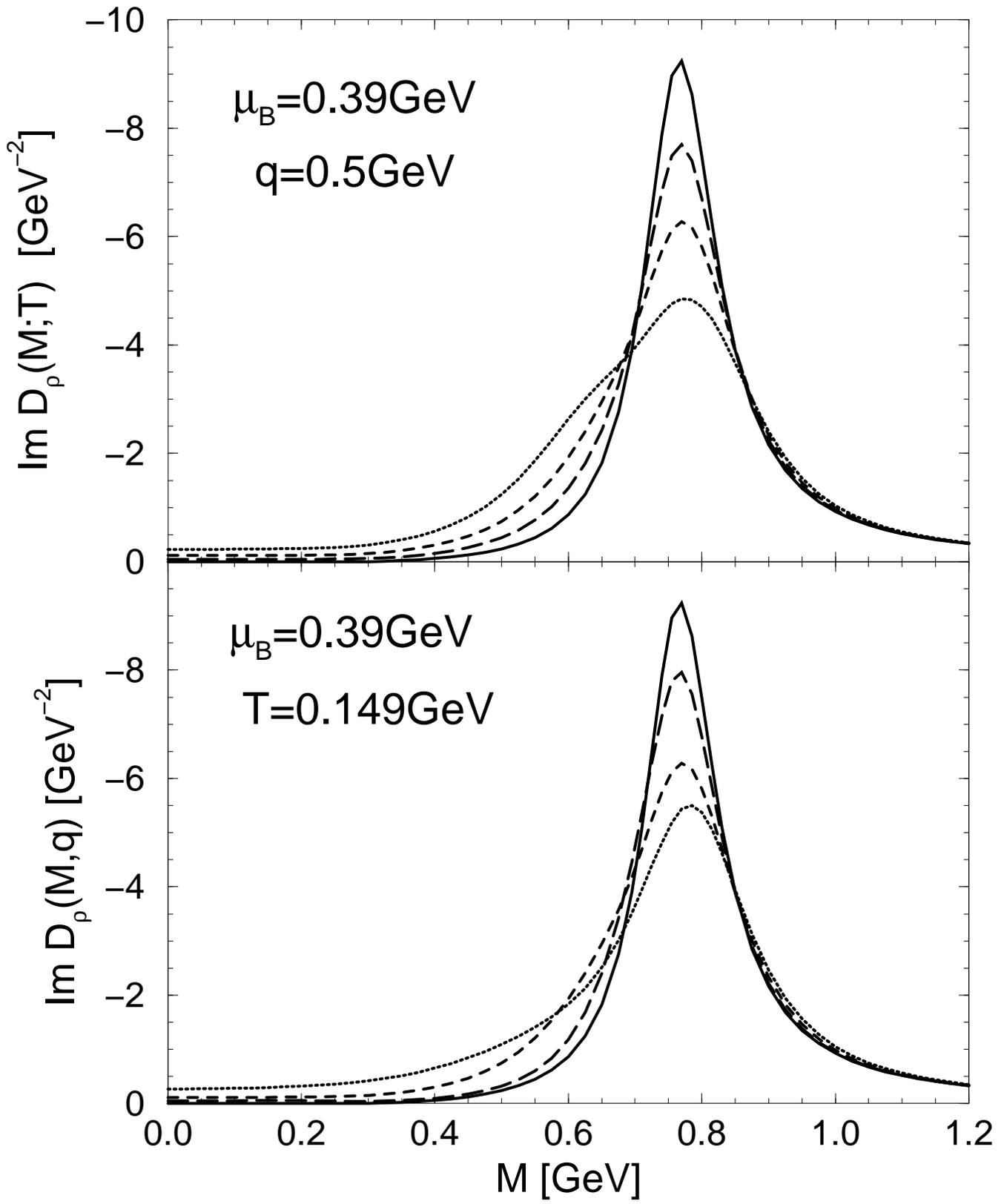
\item[{\bf Figure 5:}]
Imaginary part of the in-medium $\rho$ propagator when accounting 
for $\rho N$ and $\rho\Delta$ scattering in a hot baryon gas at 
chemical potential $\mu_B$$\equiv$$\mu_N$=$\mu_\Delta$=0.39~GeV; \\ 
upper panel: for fixed 3-momentum $q$=0.5~GeV and temperatures 
$T$=0.127~GeV (long-dashed line), $T$=0.149~GeV (short-dashed line)  
and $T$=0.170~GeV (dotted line); \\
lower panel: for fixed temperature $T$=0.149~GeV and 3-momenta 
$q$=0.25~GeV (long-dashed line), $q$=0.5~GeV (short-dashed line) and 
$q$=0.75~GeV (dotted line).  

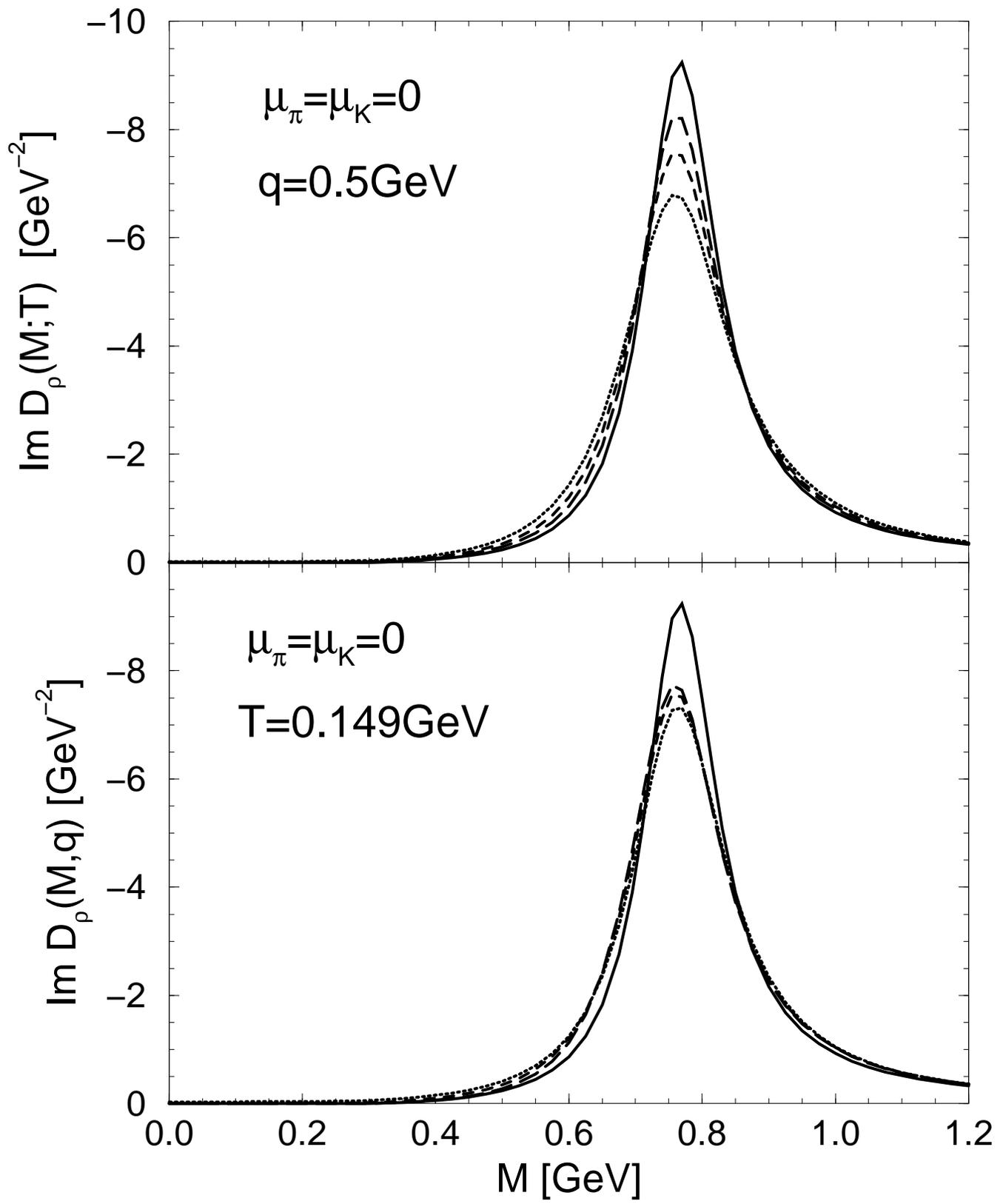
\item[{\bf Figure 6:}]
Imaginary part of the in-medium $\rho$ propagator when accounting
for $\rho\pi$ and $\rho K/\bar K$ scattering in a hot meson gas 
at zero chemical potential; \\ 
upper panel:  for fixed 3-momentum $q$=0.5~GeV and temperatures
$T$=0.127~GeV (long-dashed line), $T$=0.149~GeV (short-dashed line) 
and $T$=0.170~GeV (dotted line); \\ 
lower panel: for fixed temperature $T$=0.149~GeV and 3-momenta 
$q$=0 (long-dashed line), $q$=0.5~GeV (short-dashed line) and 
$q$=1~GeV (dotted line).  

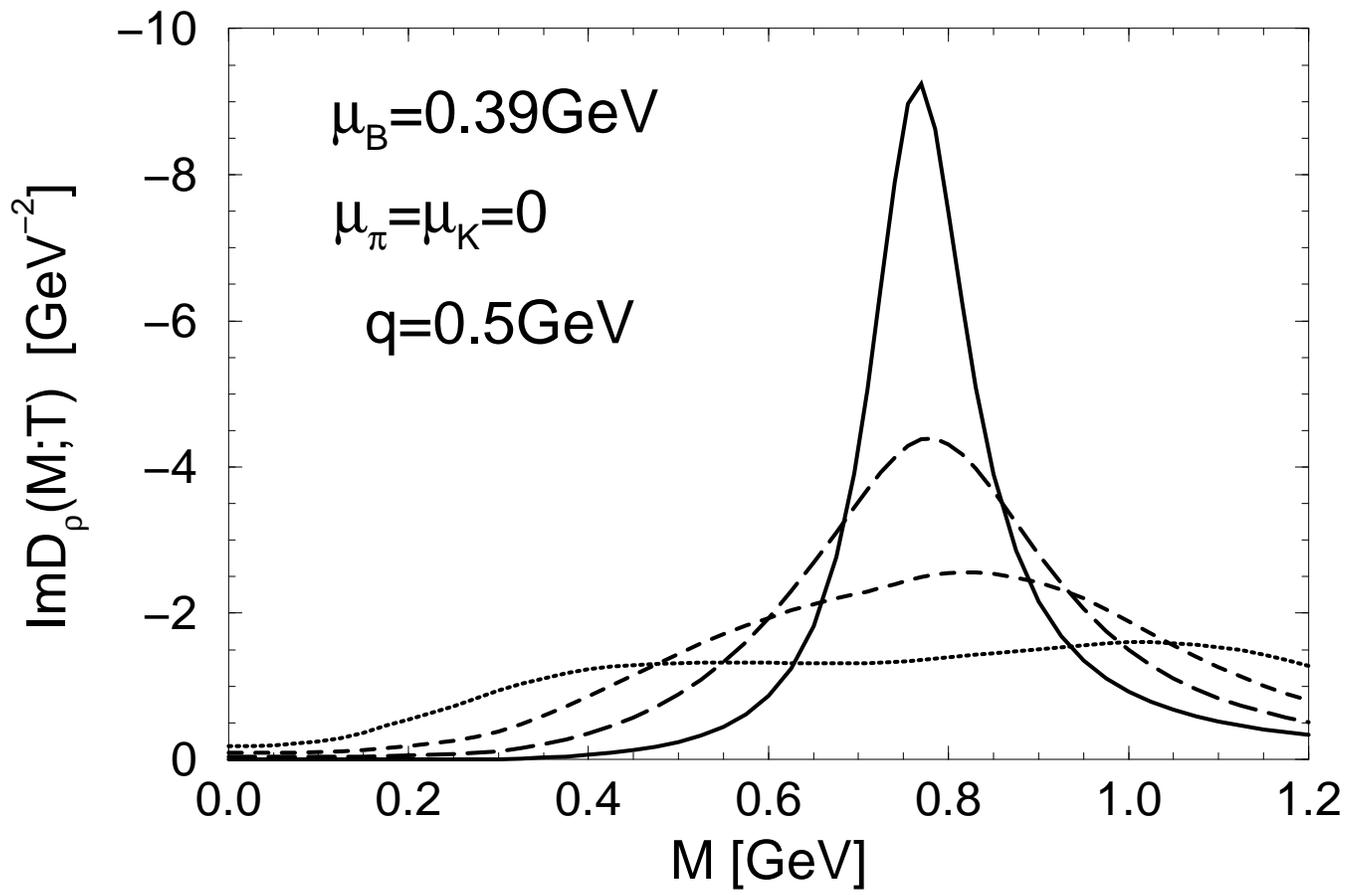
\item[{\bf Figure 7:}]
Our final results for the imaginary part of the in-medium $\rho$ 
propagator after inclusion of in-medium $\pi\pi$ states as well as 
$\rho$ meson scattering off nucleons, delta's, pions and (anti-) kaons.  
The curves correspond to fixed 3-momentum ($q$=0.5~GeV) and chemical 
potentials ($\mu_B$=0.39~GeV, $\mu_{meson}$=0) at temperatures of  
$T$=0.127~GeV (long-dashed), $T$=0.149~GeV (short-dashed) and 
$T$=0.170~GeV (dotted).  

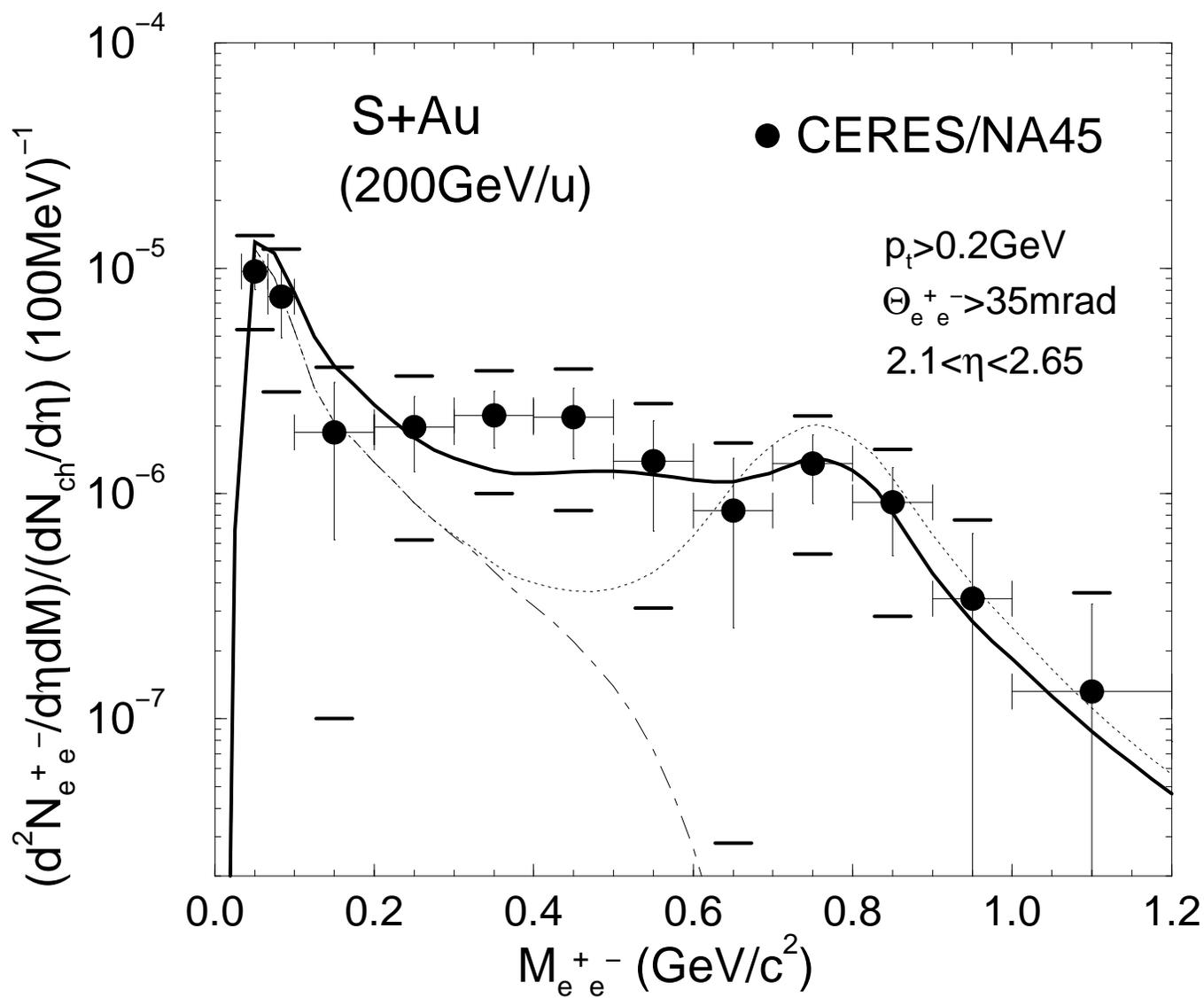
\item[{\bf Figure 8:}]
Dielectron spectrum from central S-Au (200~GeV/u) collisions; 
the dots are the CERES/NA45 data with statistical errors (vertical bars)
and systematic uncertainty added independently (resulting in the thick
horizontal bars); the dotted curve is the sum of free Dalitz 
(dashed-dotted line) and $\omega$ decays (both extracted from recent
transport results~\cite{LKB}) as well as free $\pi^+\pi^-$ 
annihilation; the full curve is obtained when evaluating the 
$\pi^+\pi^-$ contribution with the medium modified $\rho$ propagator. 
  
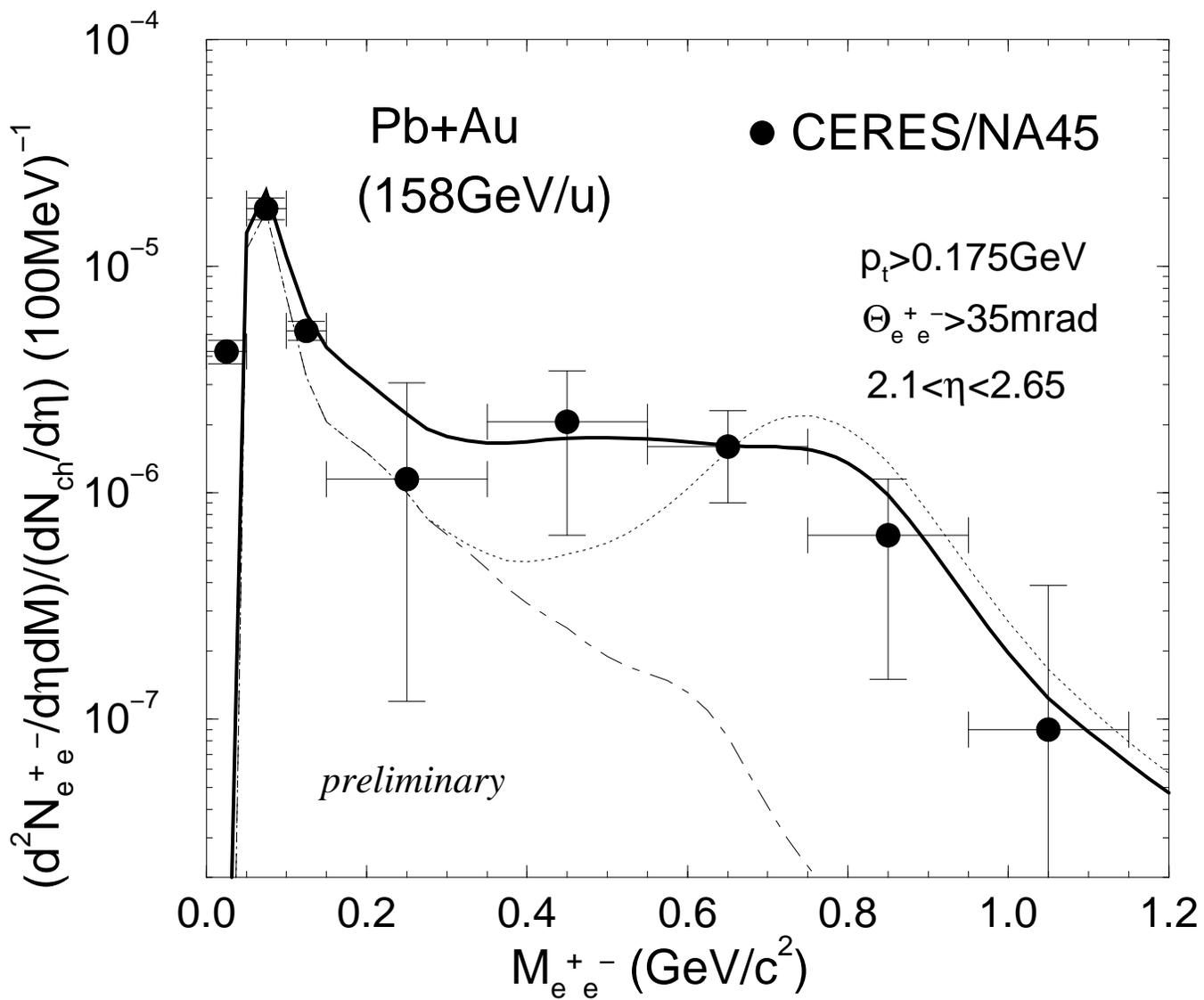
\item[{\bf Figure 9:}]
Same as fig.~8, but for central Pb-Au (158~GeV/u) collisions, where 
the dots are {\it preliminary} data from the CERES/NA45 
collaboration~\cite{Ulri}; here, only the statistical errors are 
displayed.

\end{itemize}

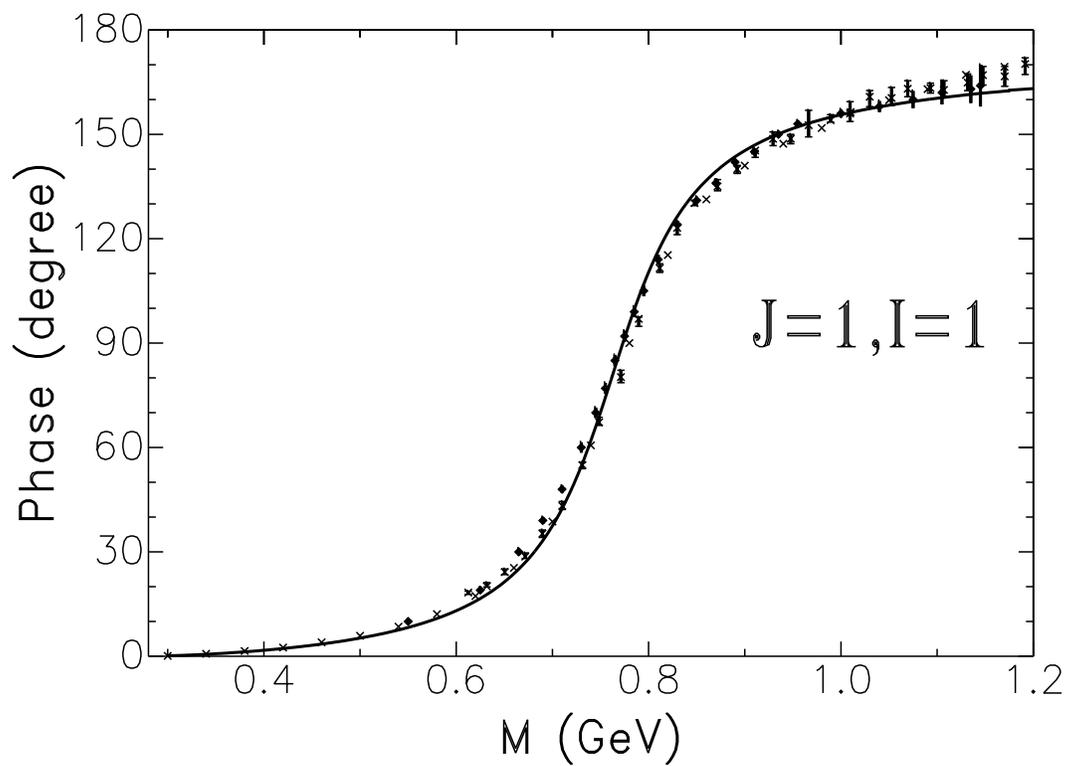
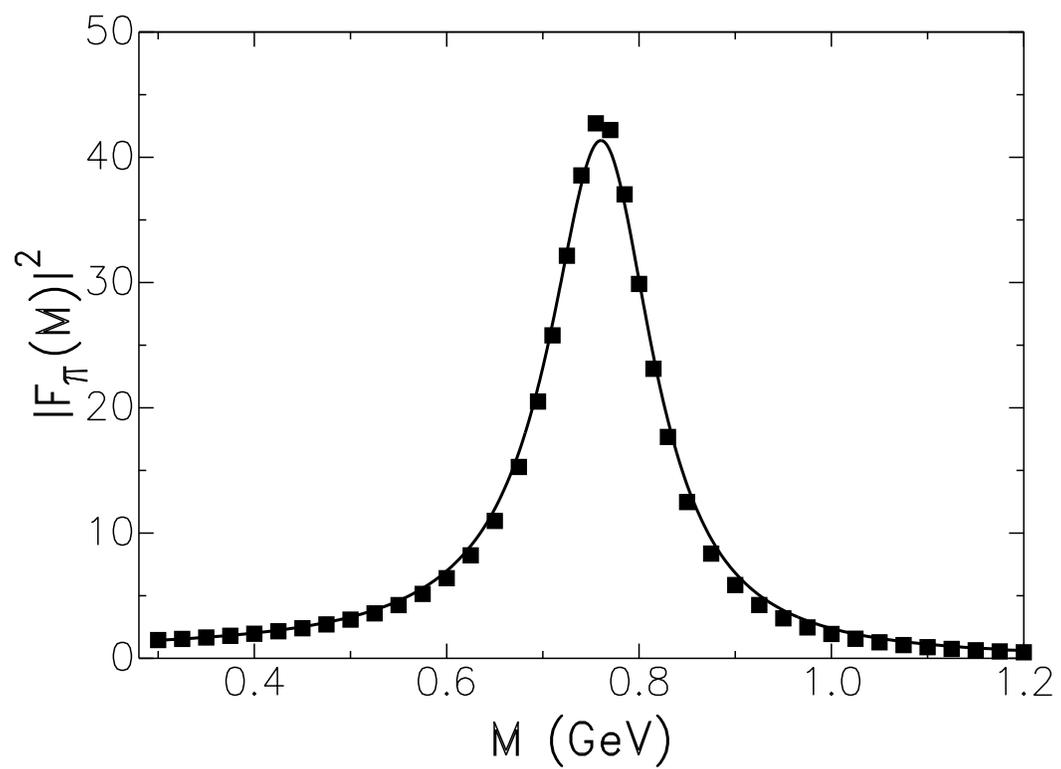
\end{document}